\documentclass[fleqn,usenatbib,useAMS]{mnras}
\usepackage{graphicx}   
\usepackage{amsmath}    
\usepackage{amssymb}    
\usepackage{multicol}        
\usepackage{bm}         
\usepackage{pdflscape}  
\usepackage[T1]{fontenc}
\usepackage{ae,aecompl}
\usepackage{newtxtext,newtxmath}

\def\kms{km s$^{-1}$}

\def\aap{A\&A}
\def\apjl{ApJ}
\def\apj{ApJ}

\def\mnras{MNRAS}
\def\nat{Nature}

\title[The Most Massive WDs]{The Most Massive White Dwarfs in the Solar Neighborhood}
\author[Kilic, Bergeron, Blouin, and B{\'e}dard]
{Mukremin Kilic$^1$, P. Bergeron$^2$, Simon Blouin$^3$, A. B{\'e}dard$^2$\\
$^1$Homer L. Dodge Department of Physics and Astronomy, University of Oklahoma, 440 W. Brooks St., Norman, OK, 73019, USA\\
$^2$D\'epartement de Physique, Universit\'e de Montr\'eal, C.P. 6128, Succ. Centre-Ville, Montr\'eal, QC H3C 3J7, Canada\\
$^3$Los Alamos National Laboratory, P.O. Box 1663,Mail Stop P365, Los Alamos, NM 87545, USA\\
}

\date{\ \ Submitted \today \vspace{-0.5cm}}
\pubyear{2021}

\begin{document}
\label{firstpage}
\pagerange{\pageref{firstpage}--\pageref{lastpage}}
\maketitle

\begin{abstract}

We present an analysis of the most massive white dwarf candidates in the Montreal White Dwarf Database 100 pc sample.
We identify 25 objects that would be more massive than $1.3~M_{\odot}$ if they had pure H atmospheres and CO cores,
including two outliers with unusually high photometric mass estimates near the Chandrasekhar limit. We provide follow-up
spectroscopy of these two white dwarfs and show that they are indeed significantly below this limit. We expand our model
calculations for CO core white dwarfs up to $M=1.334\ M_\odot$, which corresponds to the high-density limit of our
equation-of-state tables, $\rho = 10^9$ g cm$^{-3}$. We find many objects close to this maximum
mass of our CO core models. A significant fraction of ultramassive white dwarfs are predicted to form through binary mergers.
Merger populations can reveal themselves through their kinematics, magnetism, or rapid rotation rates. We identify four outliers
in transverse velocity, four likely magnetic white dwarfs (one of which is also an outlier in transverse velocity), and one with
rapid rotation, indicating that at least 8 of the 25 ultramassive white dwarfs in our sample are likely merger products.

\end{abstract}


\section{Introduction}

Stellar evolution theory tells us that stars with initial masses less than about 8 $M_{\odot}$ form degenerate cores
and evolve into white dwarfs that are supported by electron degeneracy pressure. \citet{chandrasekhar31}
showed that there is an upper mass limit for a star that is supported by electron degeneracy pressure, which is
roughly 1.4 $M_{\odot}$. \citet{takahashi13} studied the evolution of the progenitors for electron capture supernovae,
and found that when the mass interior to the helium burning shell reaches 1.367 $M_{\odot}$, the ONe core collapses due
to electron capture on $^{24}$Mg and $^{20}$Ne, leading to the formation of a neutron star \citep{miyaji80,nomoto87}.
Hence, single star evolution cannot produce white dwarfs more massive than 1.367 $M_{\odot}$.

Population synthesis models demonstrate that a significant fraction of the massive white dwarfs
above $1~M_{\odot}$ likely form through double white dwarf mergers \citep{temmink20,cheng20}.  
The mergers of two CO core white dwarfs may leave behind a single massive white dwarf
if the remnant's mass is below the Chandrasekhar limit \citep[][and references therein]{schwab21}. 

Ultramassive white dwarfs have been discovered serendipitiously in surveys of nearby white dwarfs.
GD 50 is the best example of an ultramassive white dwarf identified early on \citep{bergeron92}. 
Based on Gaia Data Release 2 photometry and astrometry \citep{gaia18}, GD 50 is an $M= 1.28 \pm 0.08 M_{\odot}$
white dwarf only 31 pc away from the Sun and likely a member of the AB Doradus moving group \citep{gagne18}.
Magnetic white dwarfs PG 1658+441 \citep{schmidt92} and RE J0317$-$853 \citep{barstow95,ferrario97}, and the nearby
DA white dwarf LHS 4033 \citep{dahn04} are similar, with mass estimates around 1.3 $M_{\odot}$.  Recent examples of
ultramassive white dwarfs include the discoveries of a 1.28 $M_{\odot}$ white dwarf in the open cluster M37 \citep{cummings16},
a rapidly rotating $\sim1.33 M_{\odot}$ DBA white dwarf \citep{pshirkov20}, and a 1.14 $M_{\odot}$ white dwarf
with a mixed hydrogen-carbon atmosphere \citep{hollands20}.

The 40 pc white dwarf sample in the northern hemisphere includes three white dwarfs with $\log{g}\geq9$ \citep{mccleery20}.
The most massive white dwarf in that sample is WD 1653+256 with $M = 1.285 \pm 0.003 M_{\odot}$.
\citet{kilic20} performed a detailed model atmosphere analysis of the 100 pc white dwarf sample in the SDSS footprint.
With a larger survey volume, they found white dwarfs with masses up to 1.36 $M_{\odot}$. J1140+2322,
with $T_{\rm eff}=11850 \pm 200$ K and $M=1.358 \pm 0.022$, is the most massive white dwarf in that sample, though
the more precise Gaia EDR3 parallax \citep{gaia20} indicates a slightly more distant and less massive white dwarf. 
Both mass estimates for WD 1653+256 and J1140+2322 assume a CO core.

\begin{figure*}
\centering
\includegraphics[width=3.2in, clip=true, trim=0.3in 0.7in 1.1in 1in]{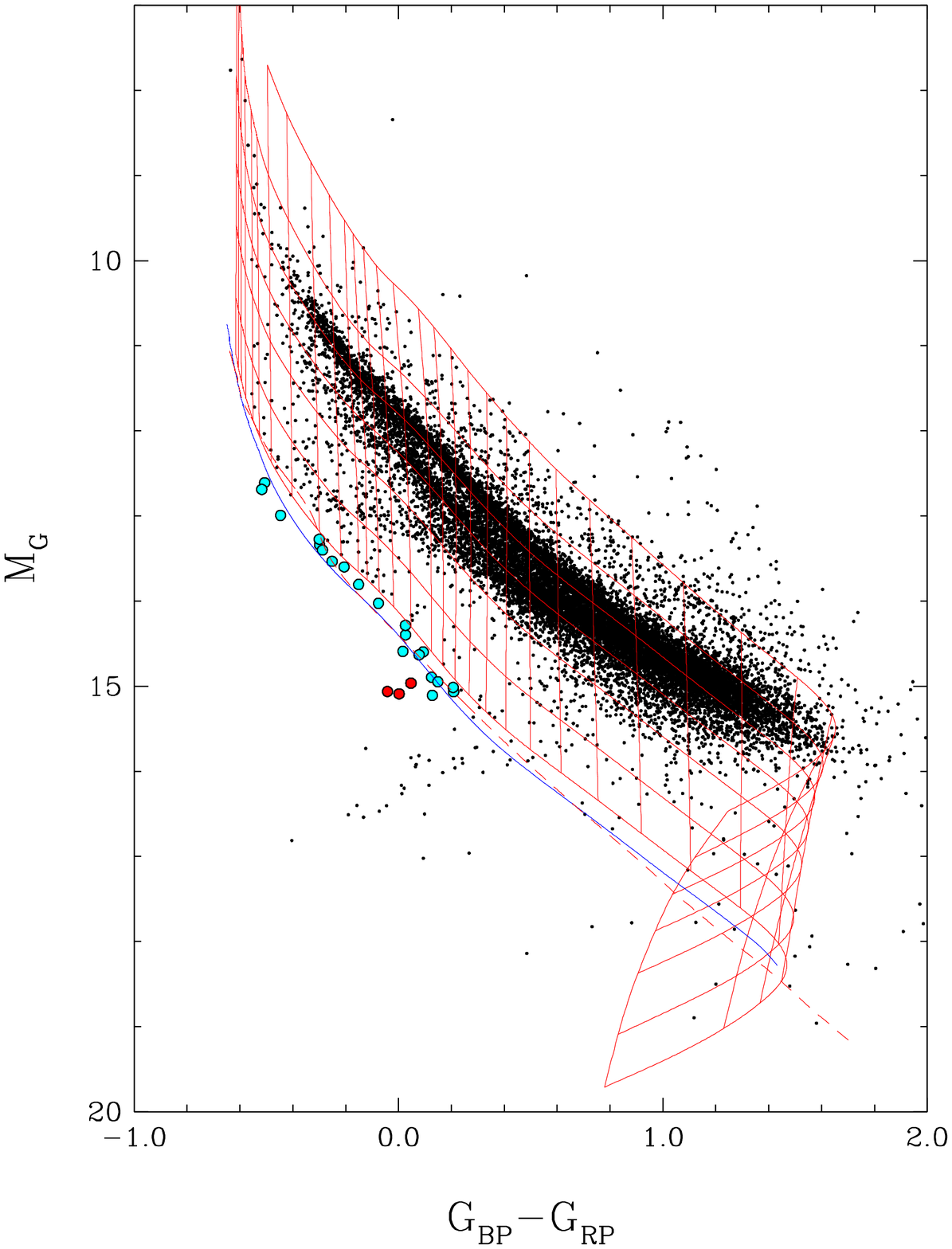}
\includegraphics[width=3.2in, clip=true, trim=0.3in 0.7in 1.1in 1in]{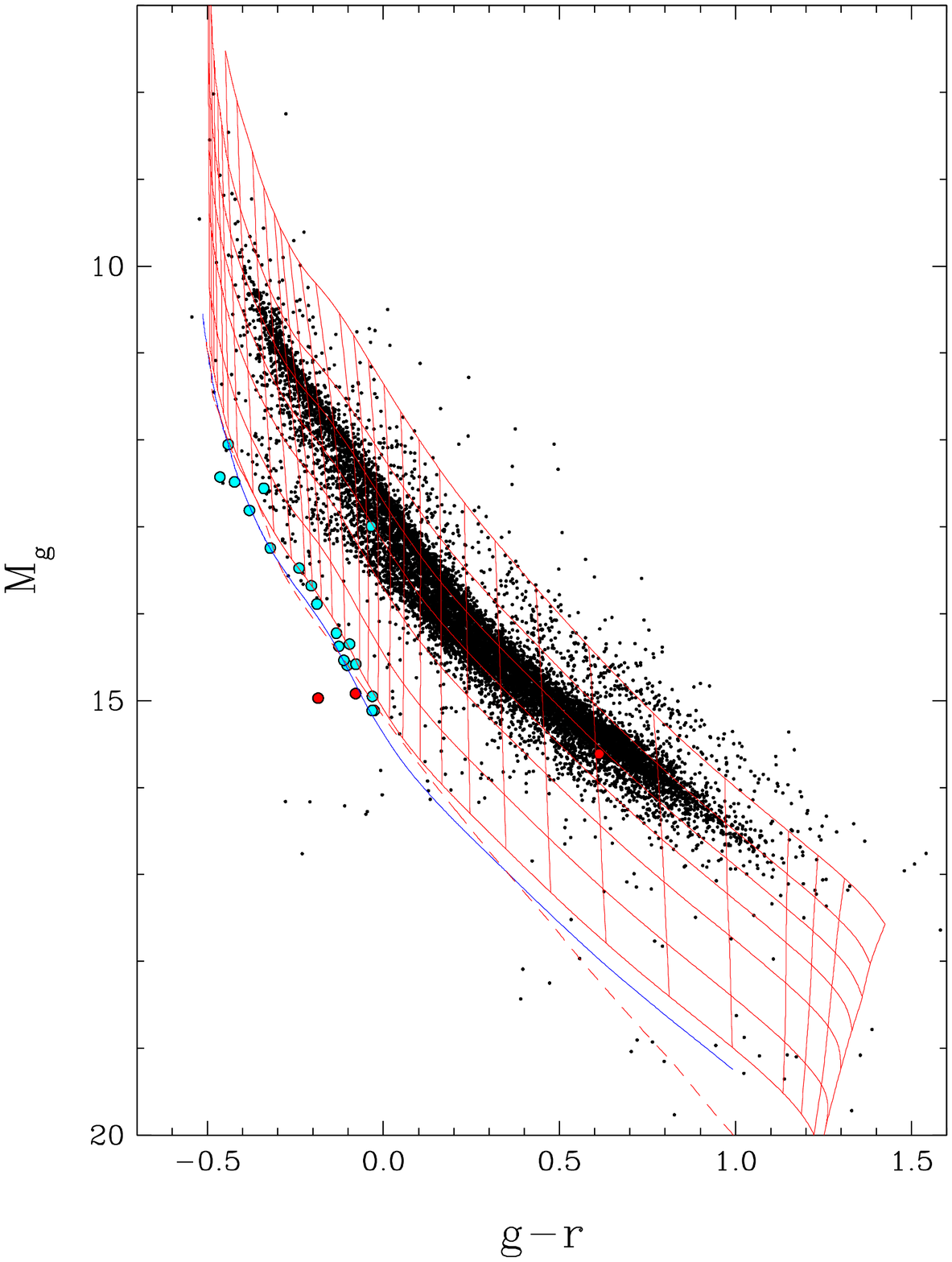}
\caption{Gaia (left) and Pan-STARRS (right) color-magnitude diagram of the 100 pc sample in the Montreal White Dwarf
Database \citep{dufour17}. Red solid lines show the cooling sequences for CO core and pure H atmosphere white dwarf
models with 0.2, 0.4, 0.6, 0.8, 1.0, 1.2, and 1.3 $M_{\odot}$ (from top to bottom). The dashed line shows the evolutionary
sequence for 1.3 $M_{\odot}$ pure He atmosphere white dwarfs, whereas the solid blue line shows the same for
1.29 $M_{\odot}$ pure H atmosphere ONe core white dwarfs. Our ultramassive white dwarf sample and the three outliers
are marked by filled cyan and red dots, respectively.}
\label{figcmd}
\end{figure*}

To search for the most massive white dwarfs in the solar neighborhood, here we present a detailed model atmosphere
analysis of the spectroscopically confirmed or candidate ultramassive white dwarfs in the Montreal White Dwarf Database 100 pc
sample \citep[MWDD,][]{dufour17}. We discuss the details of our selection of ultramassive white dwarfs, and present follow-up
spectroscopy of two of the newly identified candidates in Section 2. We provide the details of our fitting method and new evolutionary calculations in Section 3. We present the model atmosphere analysis of each object in Section 4, and discuss
the properties of the ultramassive white dwarf sample in Section 5, and conclude.

\section{Ultramassive White Dwarfs within 100 pc}

\subsection{Sample Selection}

We use the 100 pc white dwarf sample from the MWDD to select ultramassive white dwarf candidates. The MWDD
selection is based on Gaia DR2 \citep{gaia18}, and includes all candidates with $10\sigma$ significant parallax ($\varpi$),
G$_{\rm BP}$ and G$_{\rm RP}$ photometry, and $\varpi + \sigma_{\varpi}>10$ mas. To create a clean sample,
non-Gaussian outliers in color and absolute magnitude are removed using the recommendations from \citet{lindegren18},
and a cut in Gaia color and absolute magnitude is used to select the white dwarf candidates\footnote{See
http://montrealwhitedwarfdatabase.org/faq.html for details.}.

Figure \ref{figcmd} shows the color-magnitude diagram of this sample in Gaia and Pan-STARRS \citep{chambers16} filters, along with the evolutionary
sequences for 0.2, 0.4, 0.6, 0.8, 1.0, 1.2, and 1.3 $M_{\odot}$ (red lines, described further in section \ref{max}) CO core and
1.29 $M_{\odot}$ ONe core \citep[blue line,][]{camisassa19}
white dwarfs with pure H atmospheres. The dashed line shows the 1.3 $M_{\odot}$ pure He atmosphere CO core white dwarf
sequence. The main split in the white dwarf sequence due to the atmospheric
composition \citep{bergeron19}, as well as the IR-faint (also referred to as ultracool) white dwarf sequence \citep{kilic20}, blue objects with
$G_{\rm BP} - G_{\rm RP} \sim 0$ and  $M_G \sim 16$ mag, are also clearly visible in this figure. 

\begin{table*}
\centering
\caption{Ultramassive White Dwarf Candidates.}
\begin{tabular}{ccccccc}
\hline
Object  & Gaia DR2  & Gaia EDR3     & Gaia EDR3                  & Gaia EDR3                 &  Gaia DR2  & Gaia DR2 \\
            & Source ID  & $\varpi$ (mas) & $\mu$ (mas yr$^{-1}$) & $V_{\rm tan}$ (\kms)  & $G$ (mag) & $G_{\rm BP} - G_{\rm RP}$ (mag)  \\
\hline
SDSS J114012.81$+$232204.7 & 3980865789203927680 & 13.77 $\pm$ 0.29 &  62.8 & 21.6 & 18.84 & $+$0.017\\
SDSS J132926.04$+$254936.4 & 1448232907440917760 & 11.59 $\pm$ 0.10 &  23.6 &  9.7 & 17.69 & $-$0.446\\
SDSS J172736.28$+$383116.9 & 1343557102670161664 &  9.76 $\pm$ 0.31 &  49.1 & 23.9 & 19.95 & $+$0.047\\
SDSS J180001.21$+$451724.7 & 2115952197141317888 & 11.86 $\pm$ 0.10 &  72.4 & 29.0 & 18.22 & $-$0.206\\
SDSS J221141.80$+$113604.5 & 2727596187657230592 & 14.52 $\pm$ 0.33 & 172.3 & 56.2 & 19.27 & $+$0.128\\
SDSS J225513.48$+$071000.9 & 2712093451662656256 & 10.82 $\pm$ 0.37 &  45.0 & 19.7 & 19.22 & $+$0.036\\
SDSS J235232.30$-$025309.2 & 2448933731627261824 & 33.35 $\pm$ 0.08 & 698.7 & 99.3 & 17.00 & $+$0.094\\
WD J004917.14$-$252556.81  & 2345323551189913600 & 10.04 $\pm$ 0.25 &  36.2 & 17.1 & 19.07 & $-$0.076\\
WD J010338.56$-$052251.96  & 2524879812959998592 & 34.42 $\pm$ 0.10 & 176.7 & 24.3 & 17.37 & $+$0.208\\
WD J025431.45$+$301935.38  & 129352114170007680  & 10.16 $\pm$ 0.43 &  59.1 & 27.6 & 19.80 & $+$0.002\\
WD J032900.79$-$212309.24  & 5099116118775025408 & 16.98 $\pm$ 0.24 &  71.1 & 19.9 & 18.84 & $+$0.149\\
WD J042642.02$-$502555.21  & 4781653099991148928 & 12.74 $\pm$ 0.08 &  26.2 &  9.7 & 18.00 & $-$0.251\\
WD J043952.72$+$454302.81  & 253936196167057664  & 10.43 $\pm$ 0.14 &  44.9 & 20.4 & 18.27 & $-$0.288\\
WD J055631.17$+$130639.78  & 3343720447543820672 & 10.85 $\pm$ 0.53 &  58.3 & 25.5 & 19.73 & $+$0.207\\
WD J060853.60$-$451533.03  & 5567732956694899712 & 11.62 $\pm$ 0.08 &  70.5 & 28.8 & 18.00 & $-$0.299\\
WD J070753.00$+$561200.25  & 988421680189764224  & 11.52 $\pm$ 0.13 &  72.8 & 30.0 & 18.00 & $-$0.300\\
WD J080502.93$-$170216.57  & 5721057173131773184 & 22.44 $\pm$ 0.10 & 426.7 & 90.1 & 17.64 & $+$0.027\\
WD J093430.71$-$762614.48  & 5203792030921237248 & 11.85 $\pm$ 0.24 &  79.6 & 31.8 & 19.49 & $+$0.125\\
WD J095933.33$-$182824.16  & 5671878015177884032 & 16.84 $\pm$ 0.15 & 104.6 & 29.4 & 18.17 & $+$0.026\\
WD J111646.44$-$160329.42  & 3559695493657381248 & 15.52 $\pm$ 0.20 & 249.9 & 76.3 & 18.65 & $+$0.078\\
WD J125428.86$-$045227.48  & 3678497445865946624 & 11.11 $\pm$ 0.26 &   9.8 &  4.2 & 18.60 & $-$0.151\\
WD J174441.56$-$203549.05  & 4118923497232723072 & 10.05 $\pm$ 0.12 &  83.2 & 39.2 & 17.70 & $-$0.232\\
WD J181913.36$-$120856.44  & 4153618204302689920 & 19.42 $\pm$ 0.07 &  11.5 &  2.8 & 15.77 & $-$0.483\\
WD J183202.83$+$085636.24  & 4479342339285057408 & 13.24 $\pm$ 0.10 &   8.7 &  3.1 & 17.05 & $-$0.517\\
WD J190132.74$+$145807.18  & 4506869128279648512 & 24.15 $\pm$ 0.05 & 119.8 & 23.5 & 15.70 & $-$0.507\\
\hline
\end{tabular}
\end{table*}

We use these color-magnitude diagrams to select ultramassive white dwarf candidates. Masses are estimated
based on formal fits to the Gaia photometry and parallax. We restricted the candidates in absolute magnitude
($M_G < 15.2$ mag) to avoid the IR-faint white dwarf sequence.
The cyan symbols mark the candidates with $M > 1.3~M_{\odot}$ according to the full photometric
fits assuming CO core models. We identify 23 ultramassive white dwarf
candidates based on Gaia data. We repeat the same experiment using Pan-STARRS photometry
(the right panel in Figure \ref{figcmd}), and identify 22 candidates, 17 of which are common to both catalogs. Hence, we identify a total of
28 candidates for further analysis.

Strikingly, three stars in this sample (marked by red dots in Figure \ref{figcmd}) seem to be significantly below the $1.29~M_{\odot}$
sequence for ONe core white dwarfs in the Gaia color-magnitude diagram (the left panel). This is suggestive of masses significantly
above  $1.3~M_{\odot}$, and near the Chandrasekhar limit. For example, the most significant outlier in this figure,
J0254+3019 (Gaia DR2 129352114170007680) would have $M =  1.414 \pm 0.032~M_{\odot}$ based on
Gaia parallax and photometry and pure H atmosphere CO core white dwarf models. This object is also the most
significant outlier in the Pan-STARRS color-magnitude diagram, but the Pan-STARRS photometry, which provides
a broader wavelength coverage and improved constraints on the atmospheric parameters of our targets, indicates a slightly lower
mass of 1.370 $M_{\odot}$ for a pure H atmosphere white dwarf. 

The Gaia EDR3 parallax for one of these three outliers, Gaia DR2
6033039719166862336, puts it well beyond 100 pc. In addition, its $G_{\rm BP} - G_{\rm RP}$ color in Gaia DR2 is significantly
different than that in Gaia EDR3 and also the Pan-STARRS $g-r$ color. This object is marked by the red dot in the middle
of the white dwarf sequence in the Pan-STARRS color-magnitude diagram shown in the right panel of Figure \ref{figcmd}. We remove this object from the sample, as well as 
two additional objects with incomplete (1 to 3 filter) Pan-STARRS photometry, lowering our sample size to 25 objects. 

Table 1 presents the observational properties of this sample. We provide the SDSS object names,
if available. Otherwise, we provide the object names in the WD J format as reported in the MWDD. We also provide the most
recent astrometry from Gaia EDR3, but avoid using EDR3 photometry since the photometric calibration has changed
between DR2 and EDR3. Our sample includes targets with Gaia $G$ magnitudes ranging from 15.70 to 19.95 mag, with
J0254+3019 and J1727+3831, the two remaining outliers highlighted in Figure \ref{figcmd}, being the faintest.

\subsection{Observations}

We obtained follow-up optical spectroscopy of J0254+3019 and J1727+3831 using the 8m Gemini telescope equipped
with the Gemini Multi-Object Spectrograph (GMOS) as part of the queue program GN-2020A-DD-113.
We used the B600 grating and a 1$\arcsec$ slit, providing wavelength coverage from 3670 \AA\ to 6855 \AA\
and a resolution of 2 \AA\ per pixel in the $4\times4$ binned mode. We obtained two 400s exposures of J0254+3019
and two 600s exposures of J1727+3831. We used the {\sc IRAF GMOS} package to reduce these data. 

Given the lack of any absorption lines in our relatively low signal-to-noise ratio spectrum of J0254+3019, we
obtained four additional 826 s exposures as part of the program GN-2020B-FT-107. Our combined
spectra from these two Gemini programs confirm the DC spectral type for J0254+3019 and DA type for J1727+3831.

\section{Model Atmosphere Analysis}

\subsection{The Fitting Method}

We use the photometric technique as described in \citet{bergeron19}, and use the SDSS $u$ and Pan-STARRS $grizy$
photometry along with the Gaia EDR3 parallaxes to constrain the effective temperature and the solid angle, $\pi (R/D)^2$,
where $R$ is the radius of the star and $D$ is its distance. With a precise distance measurement from Gaia, we
can constrain the radius of each star directly, and therefore its mass based on the evolutionary models for a given
core composition.

The details of our fitting method, including the model grids used are further discussed in \citet{bergeron19},
\citet{genest19}, \citet{blouin19}, and \citet{kilic20}. Our model grids for both pure H and pure He atmosphere
white dwarfs extend to $\log{g} = 9.5$. We supplement the pure H and pure He atmosphere model grids with
mixed H/He atmosphere model grids with $\log{\rm H/He}=-5$ for stars with temperatures below 12,000 K.
Three of our targets are outside of the Pan-STARRS footprint; we use Gaia DR2 photometry in our analysis
for those stars. Since all of our targets are within 100 pc, we do not correct for reddening.

\subsection{The Core Composition}

The unknown core composition is the biggest uncertainty in our analysis; the mass estimates are systematically lower
for the ONe core model fits compared to the CO core fits.  
\citet{murai68} demonstrated that the transition from CO to ONe core composition in white dwarfs that evolve from
single stars occurs at $1.06~M_{\odot}$. \citet{schwab21} studied the evolution of binary white dwarf merger remnants,
and also found that the CO to ONe core transition occurs at a similar mass, $M\geq1.05~M_{\odot}$, in single white
dwarfs that formed via mergers. 

Observationally, it is impossible to constrain the core composition of a given white dwarf and rule out massive
CO core white dwarfs, unless they pulsate \citep[e.g.,][]{giammichele18}.
However, \citet{cheng19} identify a significant cooling delay in high mass white dwarfs on the Q-branch \citep{gaia18}.
They suggest $^{22}$Ne settling in CO core white dwarfs as a potential source of extra energy in these stars.
\citet{bauer20} confirm that the location of the Q branch over-density in the Gaia color-magnitude diagram is compatible
with CO cores, but not with ONe cores. This is based on the different crystallization temperatures of CO and ONe plasmas
(see their Figure 3, and also Figure 2 from \citealt{tremblay19}). Ions in an ONe plasma have a higher charge, so they interact more strongly than those in a CO plasma
and therefore freeze at a higher temperature. 
The location of the Q-branch over-density strongly favors CO cores even for white dwarfs more massive than 1.05 $M_{\odot}$. 
Based on this, we adopt the CO core solutions in this study, but also present the ONe core solutions for comparison.

\subsection{The Mass-Radius Relation}
\label{max}

\begin{figure}
\centering
\includegraphics[width=3.2in, clip=true, trim=0.3in 0.7in 1.1in 1in]{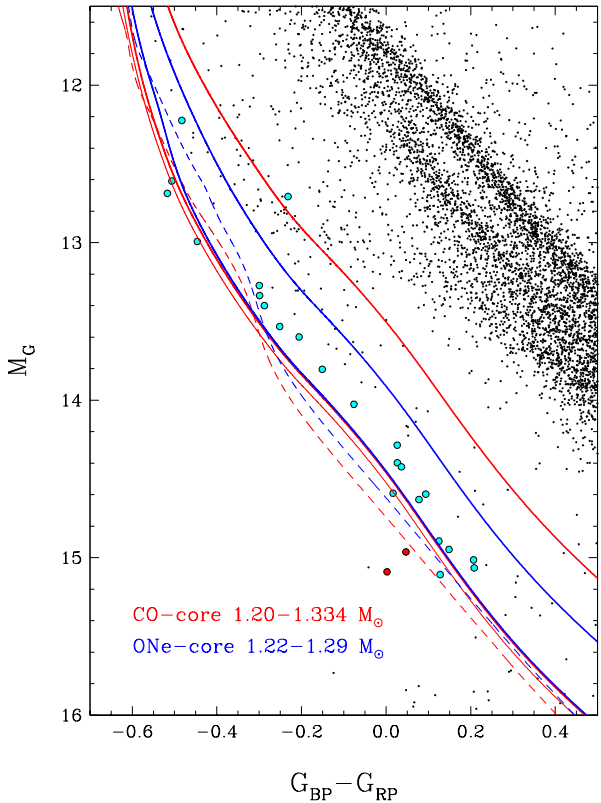}
\caption{Color-magnitude diagram of the white dwarfs in the MWDD within 100 pc (colored symbols are discussed in the text)
together with theoretical cooling sequences with CO (red) and ONe (blue) cores. The masses of the sequences are indicated
in the figure; less massive sequences are more luminous. The most massive sequences for each core composition are shown
for pure H (solid lines) and pure He (dashed lines) model atmospheres. Also displayed for the 1.334 $M_\odot$, pure H
atmospheres, CO core models, are the results for thick ($q_{\rm H}=10^{-4}$, thick red solid line) and thin
($q_{\rm H}=10^{-10}$, thin red solid line) H layers.}
\label{figzoom}
\end{figure}

\begin{figure}
\centering
\includegraphics[width=3.2in, clip=true, trim=0.2in 1.7in 0.9in 2.2in]{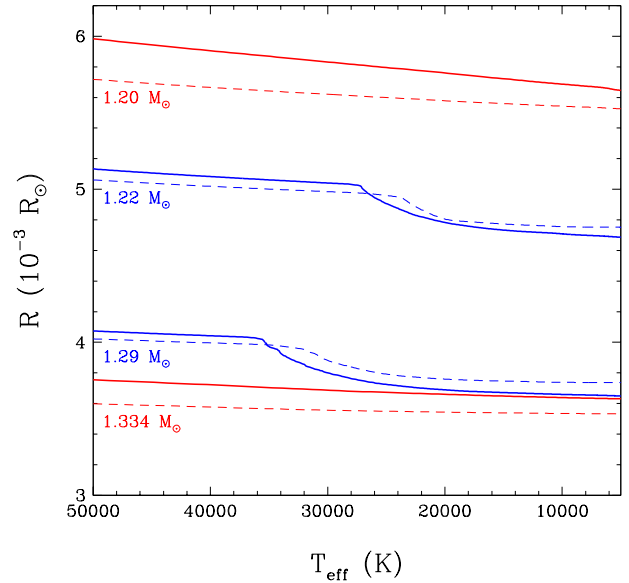}
\caption{Stellar radius as a function of effective temperature for CO core (red line) and ONe core (blue line) white dwarf models with
masses indicated in the figure.  The solid and dashed lines correspond respectively to H-rich models ($q_{\rm H}=10^{-4}$ for
CO core and $\sim$$10^{-6}$ for ONe core models, respectively) and H-poor models ($q_{\rm H}=10^{-10}$ for CO core and
no hydrogen for ONe core models, respectively).}
\label{figrt}
\end{figure}

In our fits below, we use detailed mass-radius relations of CO core
($X_{\rm C}=X_{\rm O}=0.5$) models \citep{bedard20} with masses in the
range $M=0.2-1.3\ M_\odot$ by steps of 0.05 $M_\odot$, and with thick
($q_{\rm H}\equiv M_{\rm H}/M_{\star}=10^{-4}$) and thin ($q_{\rm
  H}=10^{-10}$) H layers to fit the H- and He-atmosphere white dwarfs
in our sample, respectively. For the purpose of this analysis, we also
calculated models at $M=1.334\ M_\odot$, which corresponds to the
highest possible mass given the high-density limit of our
equation-of-state tables ($\rho = 10^9$ g cm$^{-3}$). We also rely on
the ONe core, H-rich ($q_{\rm H}\sim10^{-6}$) and H-deficient ($q_{\rm
  H}=0$) models from \citet{camisassa19} at $M=1.10$, 1.16, 1.22, and
1.29 $M_\odot$ to fit the H- and He-atmosphere white dwarfs,
respectively.

We first consider the effects of core composition, stellar mass,
surface composition, and thickness of the H layer on the predicted
colors in the color-magnitude diagram. Figure \ref{figzoom}
presents a zoomed-in version of the Gaia color-magnitude diagram previously
displayed in Figure \ref{figcmd}. Colored symbols mark our ultramassive
white dwarf sample presented in Table 1, along with the two outliers shown as red points. 
Note that one of our Pan-STARRS selected targets, J1744$-$2035, appears
redder than the $1.20~M_{\odot}$ cooling sequence based on Gaia DR2 photometry. However, improved
constraints from Pan-STARRS photometry clearly indicate that this is an
$M\sim1.3~M_{\odot}$ white dwarf (see \S \ref{utype}). 

Figure \ref{figzoom} includes cooling sequences
obtained from some of the CO and ONe core models discussed above
(shown as red and blue lines, respectively). Calculations are shown
for various masses indicated in the figure (less massive models are
more luminous). Also shown for the most massive
sequences are the results for pure H atmospheres (assuming thick H
layers; solid lines) and for pure He atmospheres (assuming thin H layers for
CO core models, or the H-deficient ONe core models from Camisassa et
al.; dashed lines); the small effect of using thick or thin H layers for pure H
atmospheres is also illustrated for the CO-core models at
$M=1.334~M_\odot$ (thick and thin red solid lines, respectively).  In
this diagram, the CO and ONe core sequences for pure H atmospheres
near $M\sim 1.2\ M_\odot$ can be used to estimate the effects of the
core composition, with the CO core models being more luminous due to
their larger radii at a given effective temperature. Interestingly
enough, the CO-core sequence at $1.334~M_\odot$ overlaps almost
perfectly with the ONe core sequence at $1.29\ M_\odot$ in this
color-magnitude diagram. However, the results for the pure He
atmospheres for the same masses, which rely on the CO-core thin H
models or the ONe core H deficient models, are quite distinct (note
that colors calculated using CO core models with no hydrogen at all
would overlap perfectly with our thin H models).

We can gain more insight into this behavior by examining the results
displayed in Figure \ref{figrt} where we show the stellar radius
as a function of effective temperature for the same models illustrated
in Figure \ref{figzoom}, for both thick and thin H layers (or H
deficient in the case of the ONe core models). For the $M\sim
1.2\ M_\odot$ models, one can see the larger radii of the CO core
models with respect to the ONe core models, as expected. The effect of
the hydrogen layer mass at high temperatures is also consistent in
both sets of models (the presence of hydrogen yields a larger radius
at a given temperature); note that our CO-core models have hydrogen
layers that are 100 times thicker than in the ONe core models. The effects
of H in our CO core models are thus more pronounced. Also
observed for the $M=1.22$ and $1.29\ M_\odot$ ONe core models is a
drop in the stellar radius due to phase separation upon
crystallization (see also Figure 11 of \citealt{camisassa19}).  This
drop in radius occurs at lower effective temperatures and with a
smaller amplitude in H deficient models. The net result is that at
lower temperatures, the H deficient ONe core models have larger radii
than their thick H counterparts, in contrast with our CO-core models
where phase separation is neglected. The effects are even more
pronounced in the most massive CO-core and ONe core models displayed
in Figure \ref{figrt}. Here the thick H layer, CO-core models at
$1.334\ M_\odot$ overlap almost perfectly with the ONe core models at
$1.29\ M_\odot$ below $T_{\rm eff}\sim30,000$~K, while the thin H (or
H deficient) models yield stellar radii that differ significantly in
the same temperature range, thus explaining the particular behavior
observed in the color-magnitude diagram displayed in Figure
\ref{figrt} for the most massive sequences.

Finally, we note that both the CO-core and ONe core models are
able to encompass most of the massive white dwarfs in our sample
displayed in Figure \ref{figzoom}, in particular if we take into
account the effect of atmospheric composition. The only exceptions are
the three red objects identified in the figure.

\section{Results}

\subsection{SDSS J172736.28$+$383116.9}

\begin{figure}
\centering
\includegraphics[width=3.2in, clip=true, trim=0.9in 3.2in 1.3in 1in]{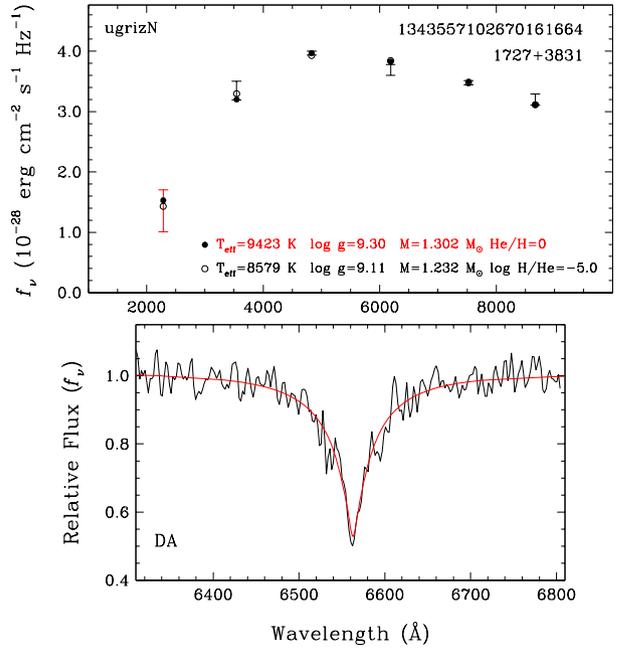}
\caption{Model fits to an ultramassive white dwarf candidate observed at Gemini. The top panel shows the best-fitting
H (filled dots) and He (open circles) atmosphere white dwarf models to the photometry (error bars), and includes the
Gaia DR2 Source ID, object name, and the photometry used in the fitting: $ugrizN$ means SDSS $u$ + Pan-STARRS
$griz$, and Galex NUV. The atmospheric parameters of the favored solution is highlighted in red. Here, and in the following
figures, we show the model parameters for CO core white dwarfs. The bottom panel shows the observed spectrum (black line)
along with the predicted spectrum (red line) based on the pure H atmosphere solution.}
\label{fig1727}
\end{figure}

Figure \ref{fig1727} shows our model fits to SDSS J172736.28$+$383116.9, hereafter J1727+3831, one of the
outliers in the Gaia color-magnitude diagram. The Pan-STARRS photometry puts this object closer to the $1.3~M_{\odot}$
models (see Figure \ref{figcmd}). The top panel shows the GALEX NUV, SDSS $u$, and Pan-STARRS
$griz$ photometry (error bars) along with the predicted fluxes from the best-fitting pure H (filled dots) and mixed
H/He (open circles) atmosphere models. The GALEX photometry is not used in the fitting, and is shown in red here and
in the following figures, but it is useful for inferring the atmospheric composition in some cases. The labels
in the same panel give the Gaia DR2 Source ID, object name, and the photometry used in the fitting. The bottom panel
shows the observed Gemini spectrum (black line) along with the predicted spectrum based on the pure H solution. Note
that we do not fit the spectroscopy data here. Instead, we simply over-plot the predicted Balmer line profile (red line)
from the photometric fit to see if a given spectrum is consistent with a pure H atmosphere composition. 

\begin{figure*}
\centering
\includegraphics[width=3.2in, trim=0.9in 3.2in 1.3in 1in]{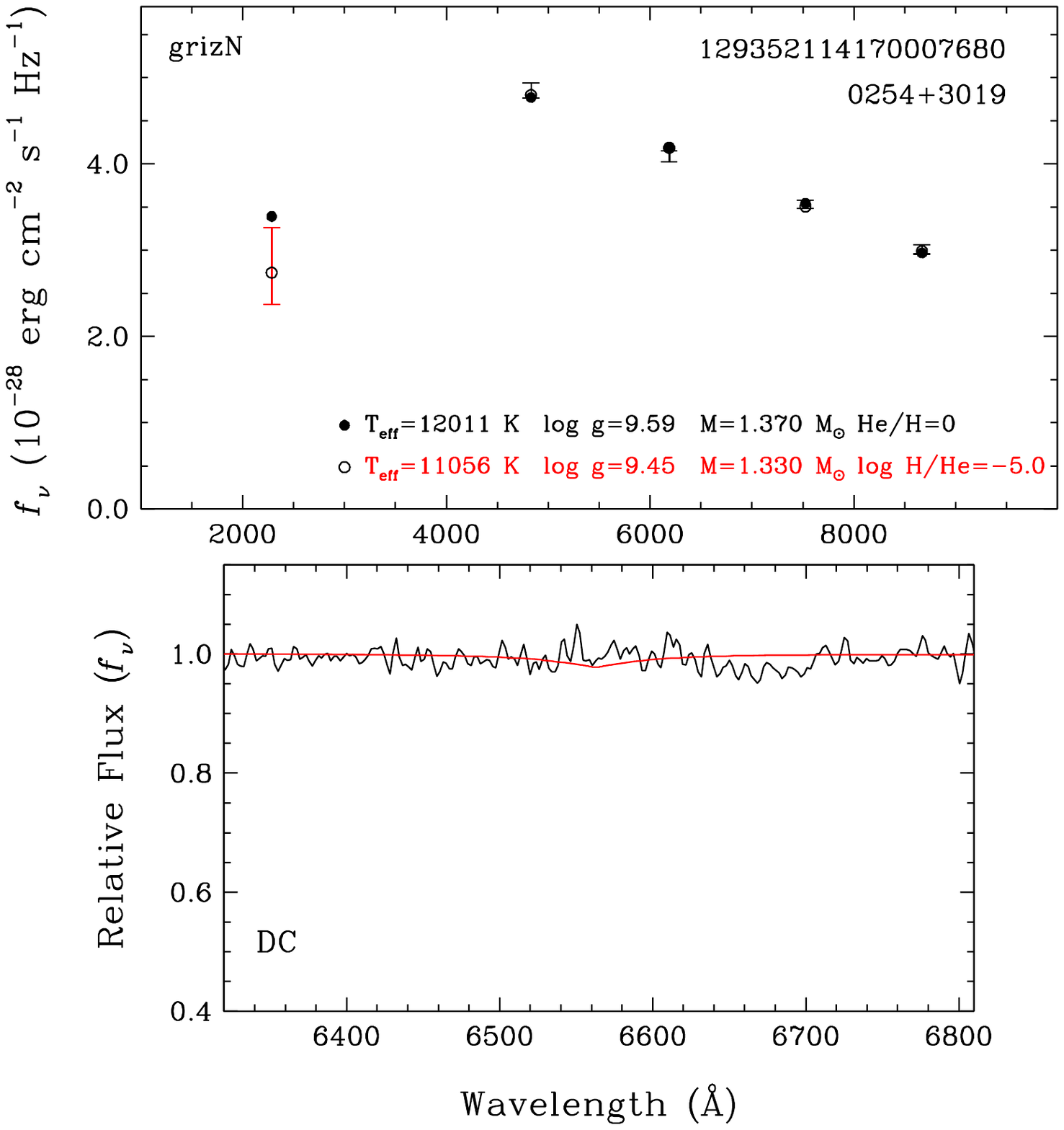}
\includegraphics[width=3.2in, trim=0.9in 3.2in 1.3in 1in]{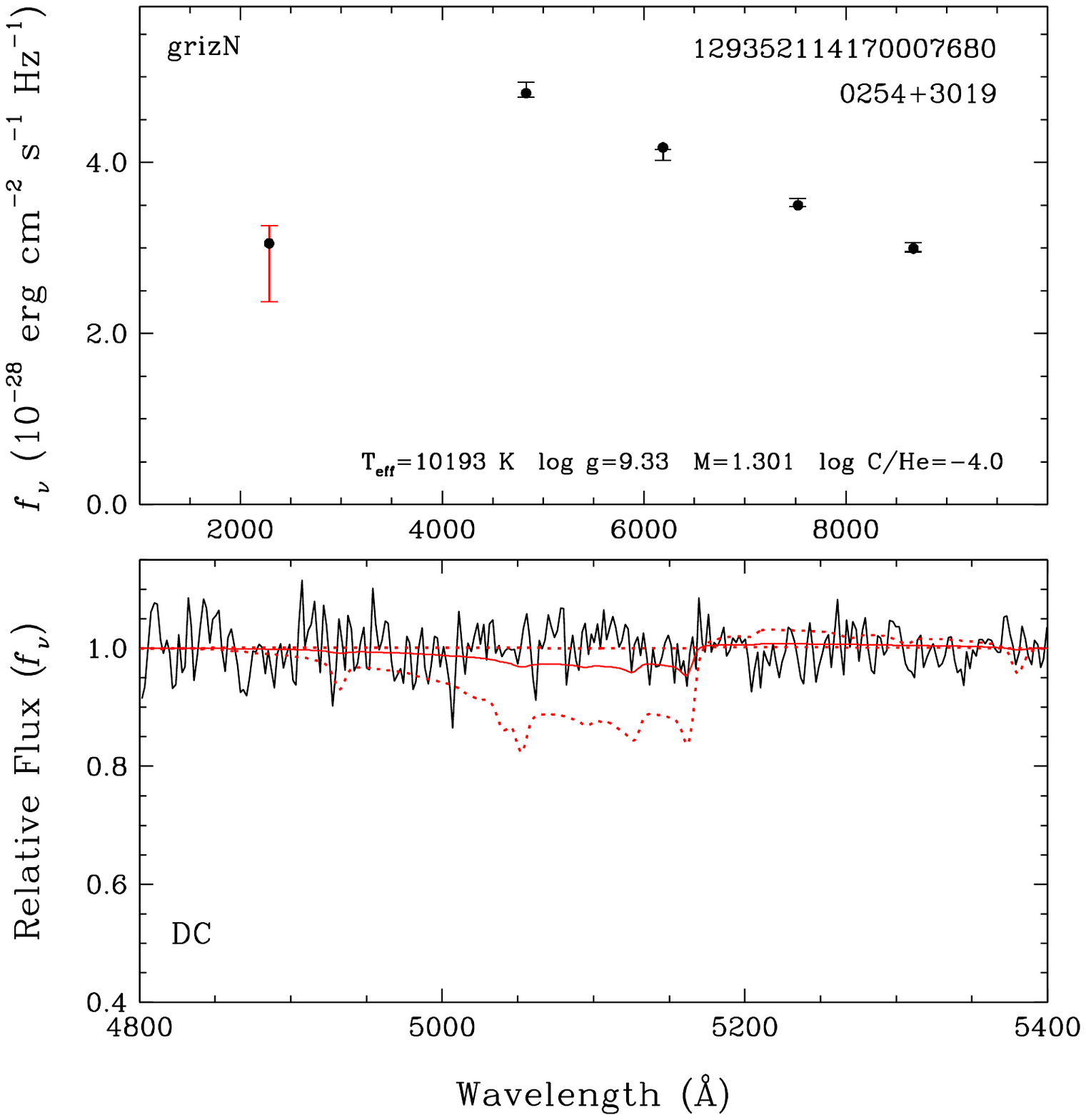}
\caption{Model fits to J0254+3019, our second target observed at Gemini. The lack of any absorption
features in its Gemini spectrum complicates our analysis of J0254+3019. The left panels show our analysis assuming pure H and
He-rich models with trace amounts of H (similar to Figure \ref{fig1727}), whereas the right panels show the results assuming
a He-rich composition with trace amounts of C. In the lower panel, the red solid line corresponds to the model spectrum obtained
from our photometric solution assuming a carbon abundance of $\log {\rm C/He}=-4$, displayed in the upper panel, while the
dotted lines show the predicted spectra assuming abundances of $\log {\rm C/He}=-3$ and $-5$. Depending on the unknown
abundances of H and C in the atmosphere, J0254+3019 has a mass of 1.30-1.33 $M_{\odot}$, assuming a CO core.}
\label{fig0254}
\end{figure*}

We cannot distinguish between the H- or He-dominated solution for J1727+3831 based on the photometry alone. However,
the observed Gemini spectrum is clearly that of a DA type star, and our photometric analysis indicates
$T_{\rm eff} = 9420 \pm 200$ K and $M = 1.302 \pm 0.011~M_{\odot}$ for a pure H atmosphere white dwarf with a CO core.
The predicted H$\alpha$ line profile for these parameters provides an excellent match to the observed spectrum.
Without additional information, there is no way to know the core composition. If J1727+3831 has an ONe core instead,
its mass would be 0.05 $M_{\odot}$ lower.

\subsection{WD J025431.45$+$301935.38}

Figure \ref{fig0254} shows our model fits to WD J025431.45$+$301935.38, hereafter J0254+3019, the most significant outlier in our
sample. Even though Pan-STARRS photometry and Gaia EDR3 parallax indicate a mass as high as 1.37 $M_{\odot}$ for
a pure H composition, it turns out that J0254+3019 is a DC white dwarf with no visible absorption features in its spectrum.
Hence, its atmosphere is clearly not dominated by H, and it appears to be an outlier in the color-magnitude diagrams
because of its atmospheric composition.

\ion{He}{I} lines disappear below about
11,000 K, which means that pure He atmosphere white dwarfs would appear as DC white dwarfs below this temperature.
However, \citet{bergeron19} demonstrated that pure He atmosphere white dwarfs are extremely rare or nonexistent in
the 6000 -- 11,000 K temperature range. Reasonable mass estimates for DC white dwarfs in this temperature range require
additional electron donors like hydrogen and carbon (or other metals). Given the unknown atmospheric composition, we explore both
possibilities for J0254+3019.

The left panels in Figure \ref{fig0254} show our model atmosphere analysis using pure H and mixed H/He models with
$\log{\rm H/He}=-5$. The top left panel shows our photometric fits, which indicate $T_{\rm eff} = 11060 \pm 560$ K and
$M = 1.330 \pm 0.016~M_{\odot}$ for a mixed H/He atmosphere and a CO core. The bottom left panel compares the predicted
H$\alpha$ line profile for this solution to the observed spectrum. We can clearly rule out H abundances greater than this limit,
as we would have detected an H$\alpha$ line. For an ONe core, the mass estimate goes down to 1.302 $M_{\odot}$.

\begin{figure}
\centering
\includegraphics[width=3.2in, trim=0.9in 3in 1in 2.5in]{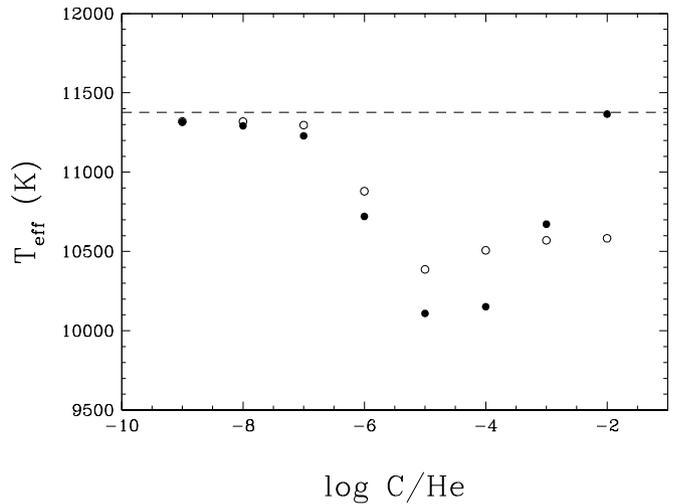}
\caption{Effective temperature determined from photometric fits to J0254+3019 as a function of the assumed carbon abundance (filled
circles). The dashed line indicates the temperature obtained from fits using pure He atmospheres. Similar results obtained using
DQ model atmospheres where the carbon line and molecular opacities have been omitted are shown by open circles.}
\label{figtc}
\end{figure}

\begin{figure*}
\centering
\includegraphics[width=2.3in, clip=true, trim=0.7in 0.5in 1in 0.8in]{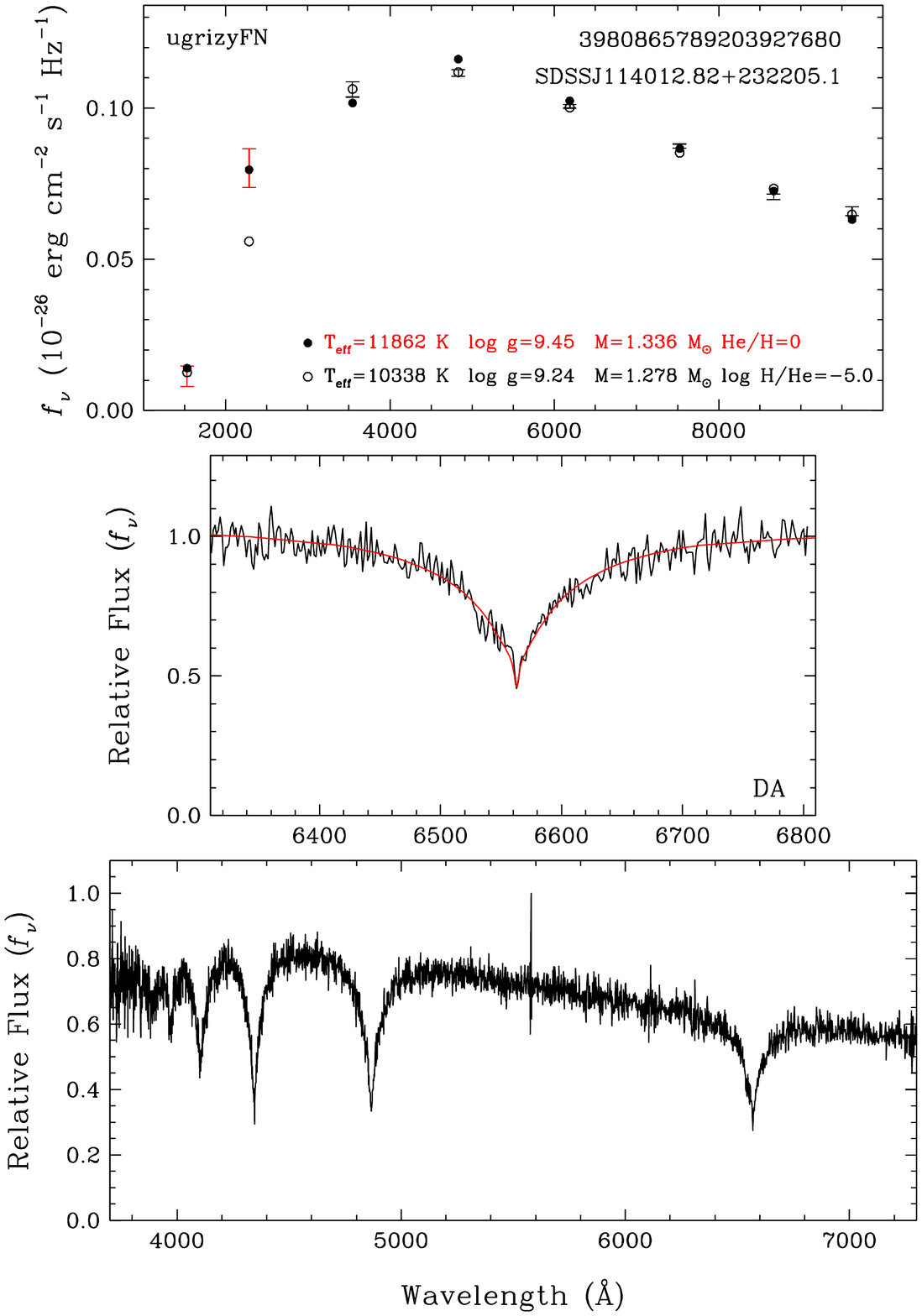}
\includegraphics[width=2.3in, clip=true, trim=0.7in 0.5in 1in 0.8in]{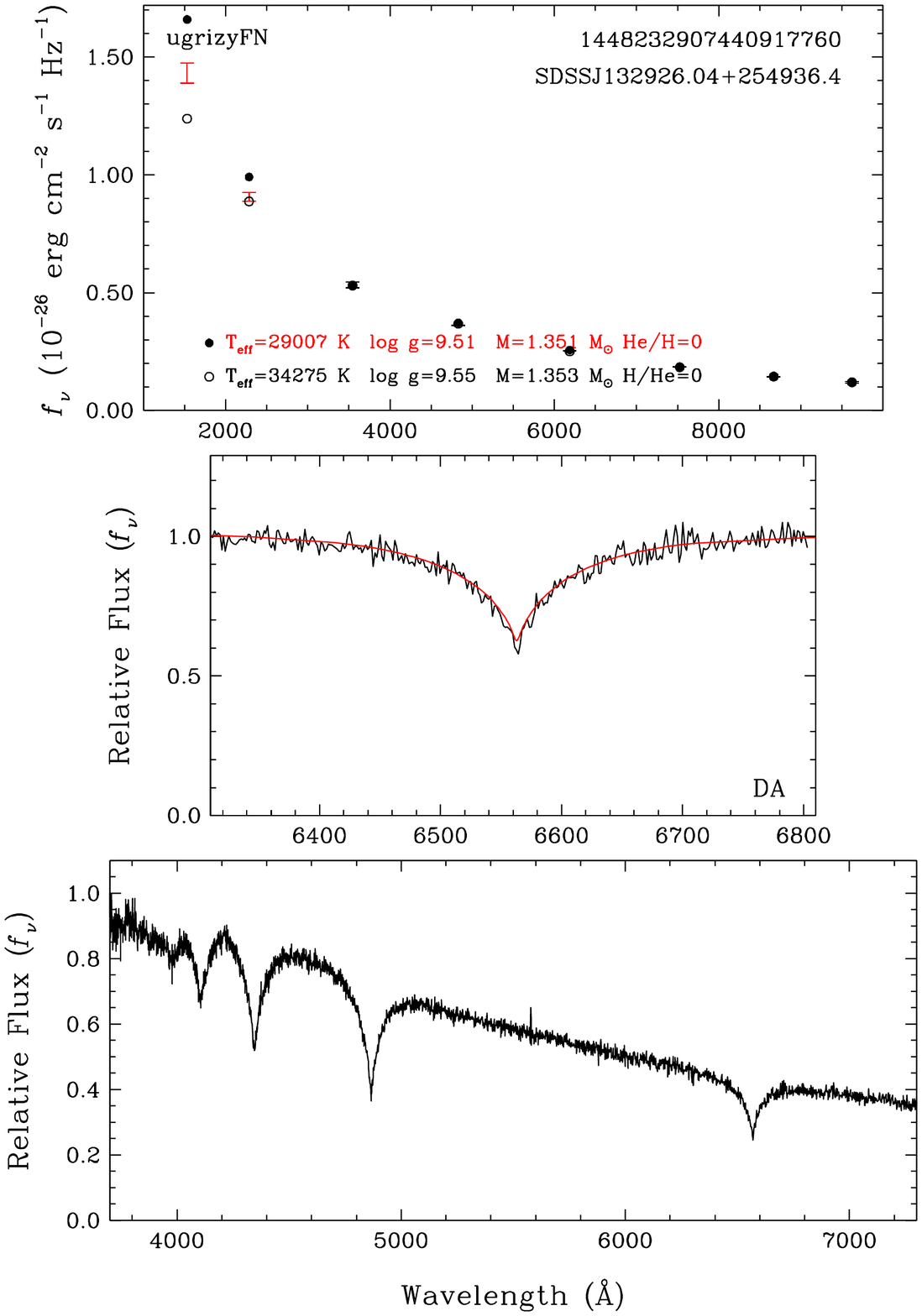}
\includegraphics[width=2.3in, clip=true, trim=0.7in 0.5in 1in 0.8in]{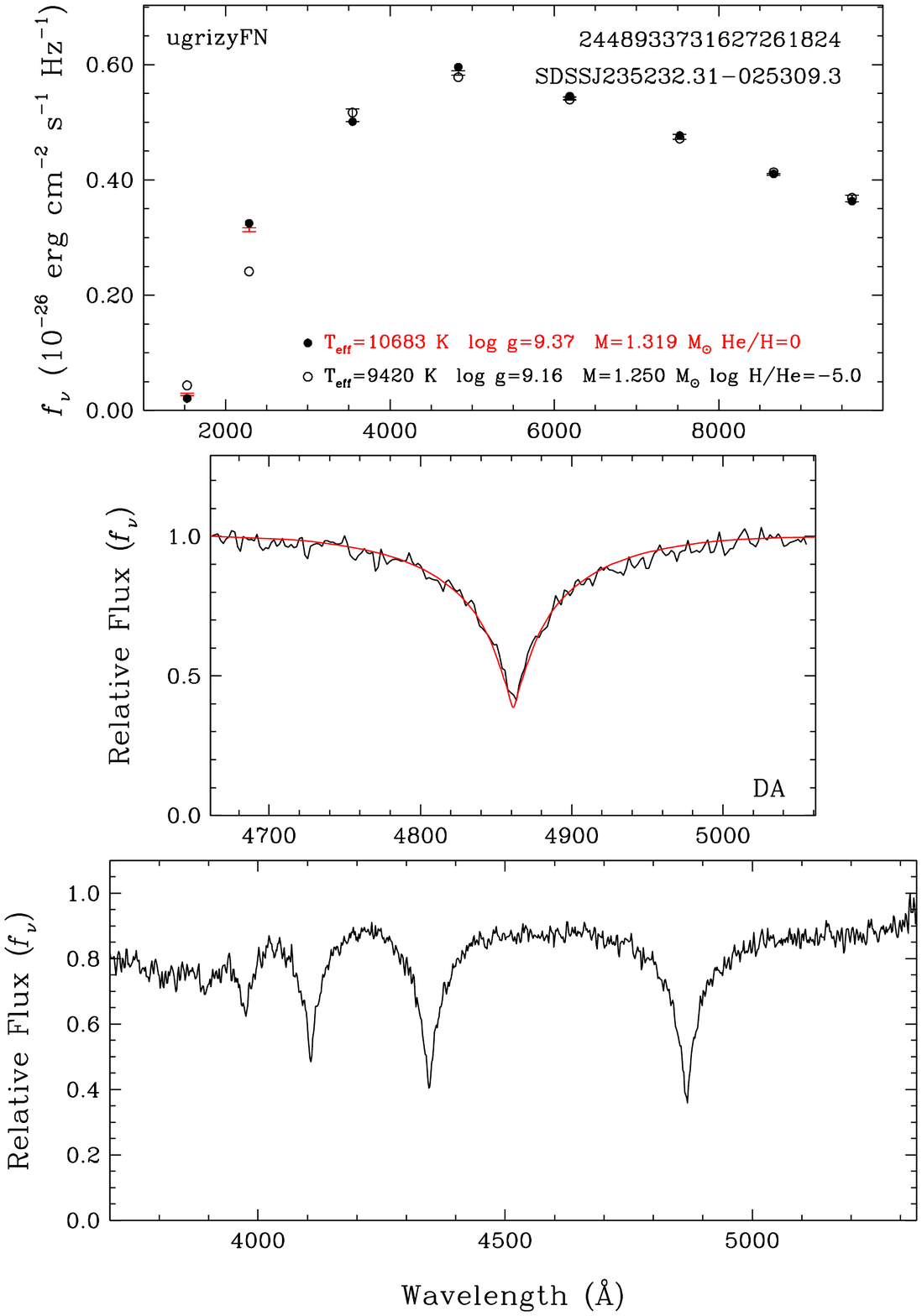}
\caption{Model atmosphere fits to three ultramassive DA white dwarfs with spectroscopy available in the literature.
The symbols  and the panels are the same as in Figure \ref{fig1727}, except that the bottom panels show a broader
wavelength range for each object.}
\label{figda}
\end{figure*}

The right panels in Figure \ref{fig0254} show the results from our analysis for a He atmosphere white dwarf with trace
amounts of carbon. As discussed by \citet{dufour05}, the physical parameters of DQ white dwarfs determined from the
photometric method are particularly sensitive to the assumed carbon abundance. In particular, Figure 8 of \citet{dufour05}
shows that the effective temperature and stellar mass can be significantly overestimated if pure He models
are used instead of models that include carbon. In this case, the presence of carbon, in addition to producing strong atomic and molecular
absorption features, also contributes to increase the number of free electrons, and thus the He$^-$ free-free opacity, particularly in the
continuum forming region, resulting in lower derived photometric temperatures.  Since larger solid angles --- $\pi(R/D)^2$ --- are required to
fit the photometric data, this implies larger derived stellar radii or smaller masses. Note that a similar reasoning can be applied in the
context of metal-rich DZ stars (see, e.g., \citealt{dufour07}),
except at very low temperatures \citep{blouin18b}.

Our photometric fits for J0254+3019, displayed in the right panels of Figure \ref{fig0254}, using DQ models assuming various carbon
abundances ($\log {\rm C/He}=-5$, $-4$, and $-3$) are consistent with the picture described above. Namely, the derived effective
temperatures and stellar masses are always smaller than the values obtained from pure He models. However, our results also revealed
that the lower the assumed carbon abundance (in the range explored here), the further away our solution was from the pure He
solution! To understand this peculiar trend better, we computed additional models with a range of carbon abundances, the results
of which are displayed in Figure \ref{figtc}. By reducing further the carbon abundance below $\log {\rm C/He}=-5$, we
gradually recover the pure He solution. The molecular Swan bands are undetectable at $\log {\rm C/He}=-5$ (see Figure
\ref{fig0254}), which implies that invisible traces of carbon can still have a large effect on the temperature structure, and thus on the
derived physical parameters.

We attempted to isolate which of the effects produced by carbon --- in terms of the equation-of-state or the opacity --- account for the
results displayed in Figure \ref{figtc}. For instance, we also show in Figure \ref{figtc} (open circles) the results obtained from
model atmospheres of DQ stars where the carbon line and molecular opacities have been turned off. While the difference in temperature
between the $\log {\rm C/He}=-5$ and $-2$ solutions has been significantly reduced, the derived photometric temperatures still show
a non monotonous variation as a function of the carbon abundance, indicating a complex interplay between the effects produced by carbon,
both in terms of the equation-of-state and the opacity calculations.

The strongest carbon features are expected in the 5000-5200 \AA\ range, and the lack of any significant
features in our Gemini spectrum limits the carbon abundance to $\log{\rm C/He}\leq-4$. If J0254+3019
has trace amounts of carbon in its atmosphere, then the best-fitting mass would be $1.301 \pm 0.014~M_{\odot}$ for a CO core,
and $1.261 \pm  0.016~M_{\odot}$ for an ONe core. Hence, depending on the unknown atmospheric and core composition, J0254+3019
has a mass in the range 1.26 -- 1.33 $M_{\odot}$.

\citet{coutu19} presented an analysis of the DQ white dwarfs in the MWDD, and showed that massive DQ white dwarfs
hotter than 10,000 K have carbon abundances of $\log{\rm C/He}>-4$. We can safely rule out such a carbon abundance
in J0254+3019, and it is unlikely that J0254+3019 belongs to the massive DQ white dwarf population. Hence, the mixed H/He atmosphere
solution discussed above is probably more representative of the physical parameters of this star.

\begin{table*}
\centering
\caption{Physical Parameters of the Spectroscopy Sample assuming ONe or CO cores. All solutions
above 1.29 $M_{\odot}$ for ONe core models and above 1.334 $M_{\odot}$ for CO core models are extrapolated.}
\begin{tabular}{cccccccc}
\hline
              &                     &                &                       &       ONe core & ONe  core & CO core & CO core\\
Object  & Composition & Spectral & $T_{\rm eff}$ &  Mass &  Cooling Age & Mass & Cooling Age\\
        &             & Type     & (K)           & ($M_{\odot}$) &   (Gyr)   &  ($M_{\odot}$) & (Gyr) \\
\hline
J010338.56$-$052251.96 & H & DAH: & 9040 $\pm$ 70 & 1.262 $\pm$ 0.003 & 2.84 $\pm$ 0.03 & 1.310 $\pm$ 0.003 & 2.60 $\pm$ 0.04\\
J025431.45+301935.38 & $\log{\rm H/He}=-5$ & DC & 11060 $\pm$ 560 & 1.302 $\pm$ 0.024 & 2.25 $\pm$ 0.10 & 1.330 $\pm$ 0.016 & 1.49 $\pm$ 0.17\\
\dots                            & $\log{\rm C/He=-4}$ & DC & 10190 $\pm$ 290 & 1.261 $\pm$ 0.016 & 2.53 $\pm$ 0.08  & 1.301 $\pm$ 0.014 & 1.93 $\pm$ 0.12\\
J114012.81+232204.7 & H & DA & 11860 $\pm$ 220 & 1.294 $\pm$ 0.008 & 2.10 $\pm$ 0.04 & 1.336 $\pm$ 0.006 & 1.71 $\pm$ 0.06 \\
J132926.04+254936.4 & H & DA & 29010 $\pm$ 750 & 1.314 $\pm$ 0.006 & 0.81 $\pm$ 0.05 & 1.351 $\pm$ 0.006 & 0.37 $\pm$ 0.03 \\
J172736.28+383116.9 & H & DA & 9420 $\pm$ 200 & 1.252 $\pm$ 0.012 & 2.78 $\pm$ 0.08  & 1.302 $\pm$ 0.011 & 2.59 $\pm$ 0.12 \\
J183202.83+085636.24 & He & DBA & 34210 $\pm$ 1020 & 1.301 $\pm$ 0.006 & 0.45 $\pm$ 0.03 & 1.319 $\pm$ 0.004 & 0.20 $\pm$ 0.02 \\
J190132.74+145807.18 & H & DC & 29100 $\pm$ 480 & 1.279 $\pm$ 0.003 & 0.61 $\pm$ 0.02 & 1.319 $\pm$ 0.004 & 0.35 $\pm$ 0.02 \\
\dots                             & He & DC & 37070 $\pm$ 720 & 1.318 $\pm$ 0.004 & 0.40 $\pm$ 0.01 & 1.330 $\pm$ 0.003 & 0.14 $\pm$ 0.01 \\
J221141.80+113604.5 & H & DAH & 9020 $\pm$ 160 & 1.262 $\pm$ 0.009 & 2.85 $\pm$ 0.07 & 1.310 $\pm$ 0.008 & 2.61 $\pm$  0.11\\
J225513.48+071000.9 & H & DC & 10990 $\pm$ 210 & 1.252 $\pm$ 0.012 & 2.36 $\pm$ 0.05 & 1.302 $\pm$ 0.011 & 2.18 $\pm$ 0.09\\
\dots                            & $\log{\rm H/He}=-5$ & DC & 9530 $\pm$ 170 & 1.188 $\pm$ 0.014 & 2.88 $\pm$ 0.06 & 1.216 $\pm$ 0.018 & 2.48 $\pm$ 0.07\\
J235232.30$-$025309.2 & H & DA & 10680 $\pm$ 100 & 1.272 $\pm$ 0.003 & 2.38 $\pm$ 0.02 & 1.319 $\pm$ 0.003 & 2.10 $\pm$ 0.03 \\
\hline
\end{tabular}
\end{table*}

\begin{figure*}
\centering
\includegraphics[width=2.3in, clip=true, trim=0.7in 0.5in 1in 0.8in]{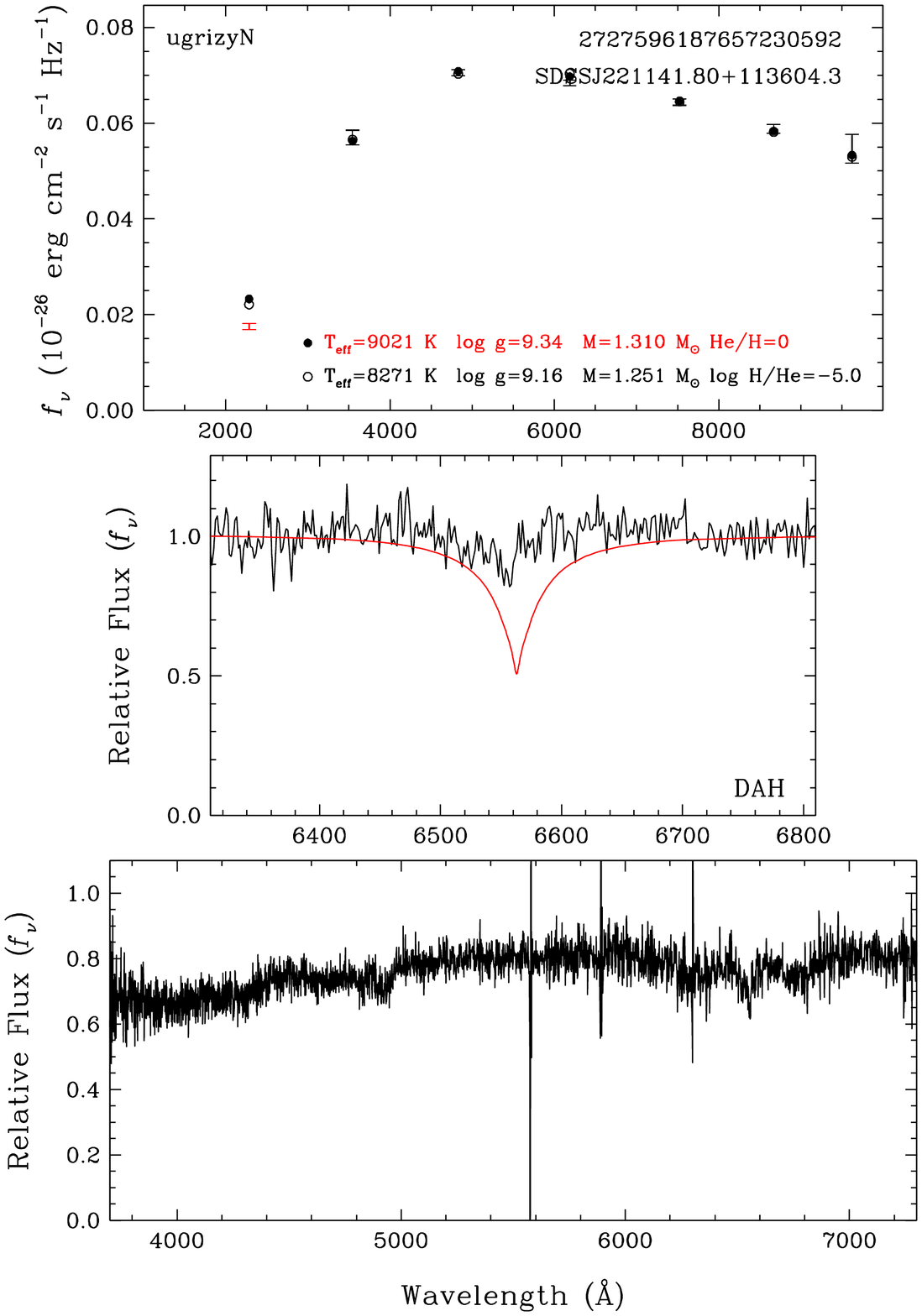}
\includegraphics[width=2.3in, clip=true, trim=0.7in 0.5in 1in 0.8in]{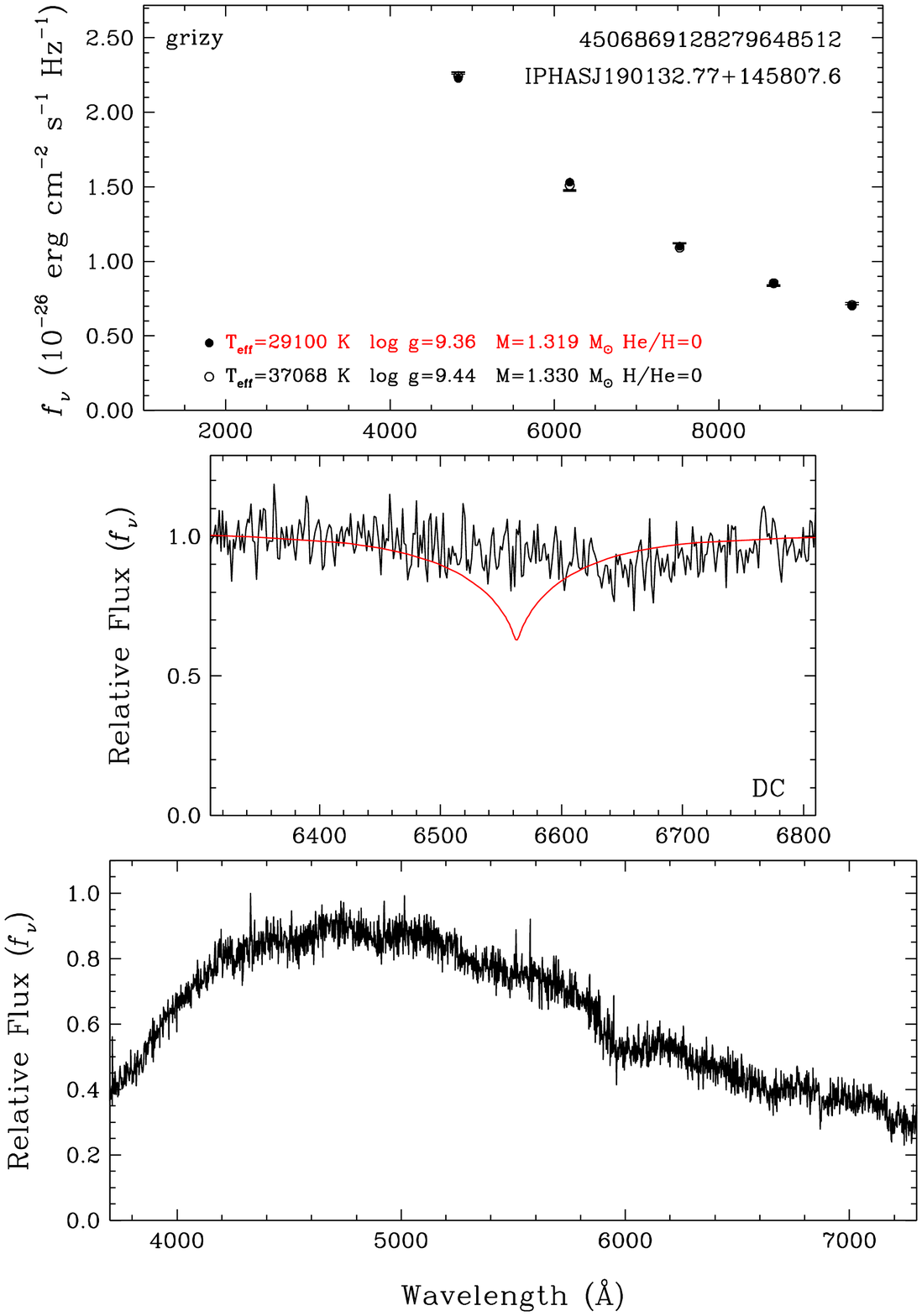}
\includegraphics[width=2.3in, clip=true, trim=0.7in 0.5in 1in 0.8in]{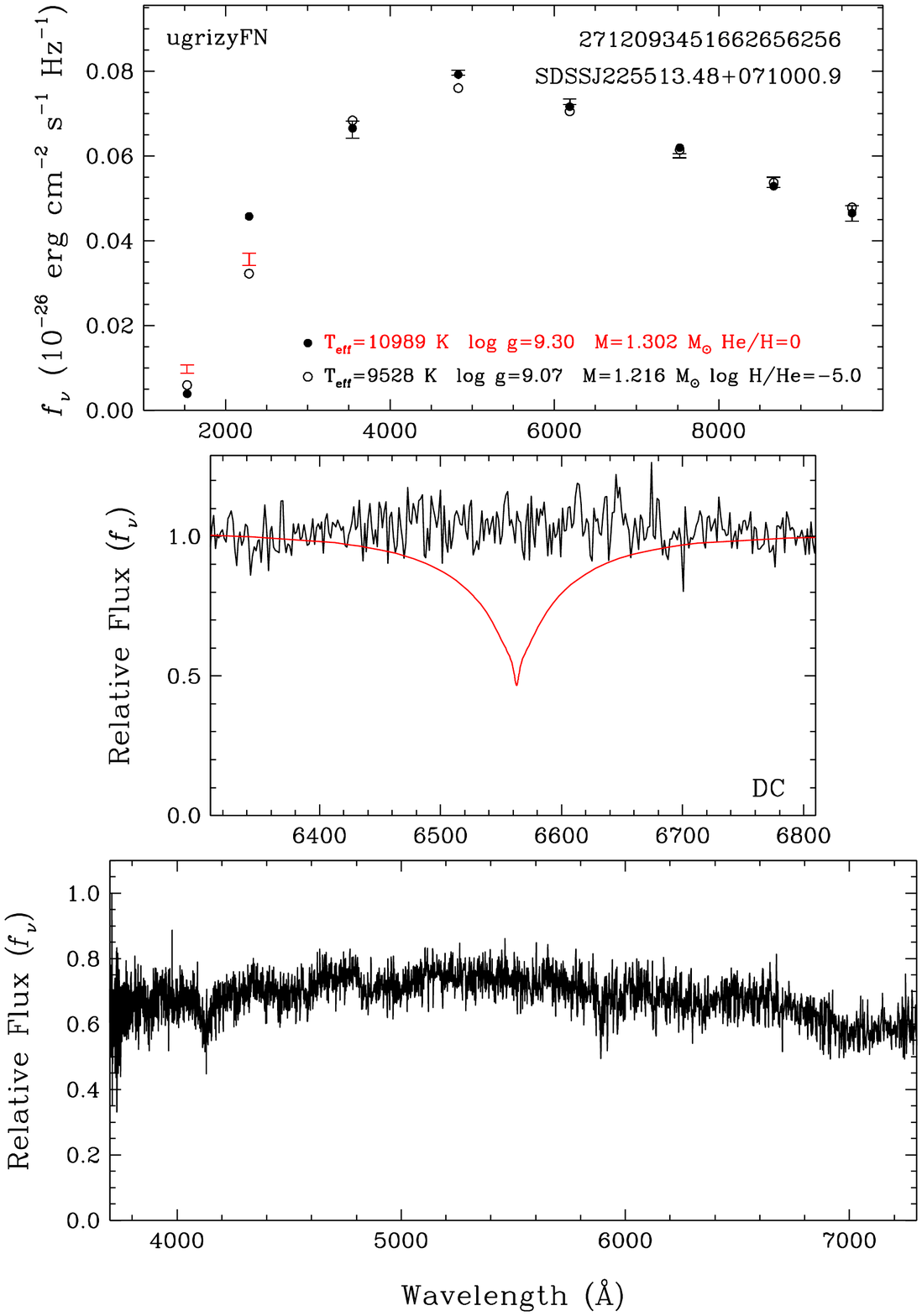}
\caption{Model atmosphere fits to three magnetic white dwarf candidates with spectroscopy available in the literature. The symbols 
and the panels are the same as in Figure \ref{figda}. Note that J1901+1458 spectrum is not flux-calibrated.}
\label{figdc}
\end{figure*}

\subsection{Additional Targets with Follow-up Optical Spectroscopy}

\subsubsection{DA White Dwarfs}

There are eight additional targets in our sample with follow-up spectroscopy available in the literature: four have
spectra in the SDSS, and four others have spectra presented in \citet[IPHAS J190132.77+145807.6]{deacon09},
\citet[LHS 4033]{gianninas11}, \citet[G270-126]{tremblay20}, and \citet[J1832+0856]{pshirkov20}.
These include 5 DA white dwarfs, two of which are magnetic, 1 DBA, and 2 DC white dwarfs. 

Figure \ref{figda} shows our model atmosphere analysis for the three non-magnetic DAs in our sample. The symbols
and the panels are the same as in Figure \ref{fig1727}, but we show an additional panel at the bottom to display a broader
wavelength range for each object. These three stars have the entire set of Galex FUV and NUV, SDSS $u$,
and Pan-STARRS $grizy$ photometry available, which enable precise constraints on their temperatures, radii, and therefore
masses. The photometric solutions provide an excellent match to the observed H line profiles, demonstrating
that they have pure H atmospheres.

J2352$-$0253 (LHS 4033, the right panels) is one of the best studied ultramassive white dwarfs in the literature.
\citet{dahn04} used both the photometric and spectroscopic method to demonstrate that this is an extremely massive
white dwarf. They used the photometric technique with a ground-based parallax measurement from the USNO to derive 
$T_{\rm eff} = 10,900 \pm 290$ K and $M=1.31-1.33~M_{\odot}$, depending on the core composition. \citet{dahn04} had
access to evolutionary models up to only $1.2~M_{\odot}$, and they adopted the \citet{hamada61} mass-radius relation for both the
spectroscopic and photometric techniques. Hence, our analysis supersedes the results presented there.
Nevertheless, our estimates of $T_{\rm eff} = 10680 \pm 100$ K and $M= 1.319 \pm 0.003~M_{\odot}$ under the assumption of
a CO core are entirely consistent with the results from \citet{dahn04}. 

The other DA white dwarfs in this figure, J1140+2322 and J1329+2549, are both hotter and more massive.
The best-fitting mass estimates for CO cores are $M=1.336 \pm 0.006~M_{\odot}$ and $1.351 \pm 0.006~M_{\odot}$, respectively.
Both of these measurements are above the highest mass CO core model available (1.334 $M_{\odot}$), and thus they
are extrapolated and should be used with caution.

Table 2 presents the physical parameters of all 10 ultramassive white dwarfs with follow-up spectroscopy available, including these three DA white dwarfs.
The best-fitting temperatures do not depend on the core composition, but the masses and the cooling ages do. For completeness,
we provide the masses and the cooling ages for both ONe and CO core compositions in Table 2.

\begin{figure*}
\centering
\includegraphics[width=2.3in, clip=true, trim=0.9in 6.3in 1in 1.0in]{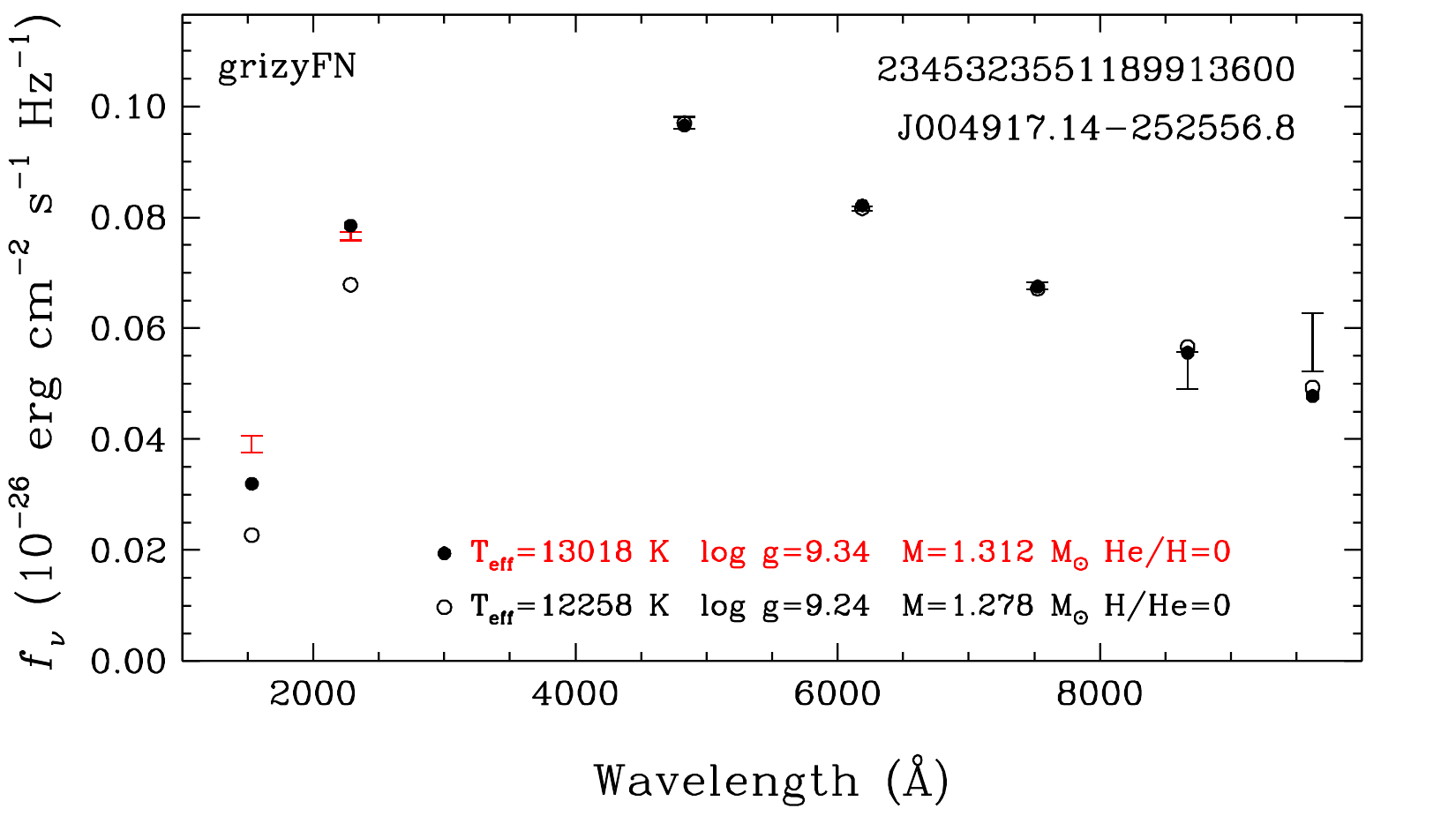}
\includegraphics[width=2.3in, clip=true, trim=0.9in 6.3in 1in 1.0in]{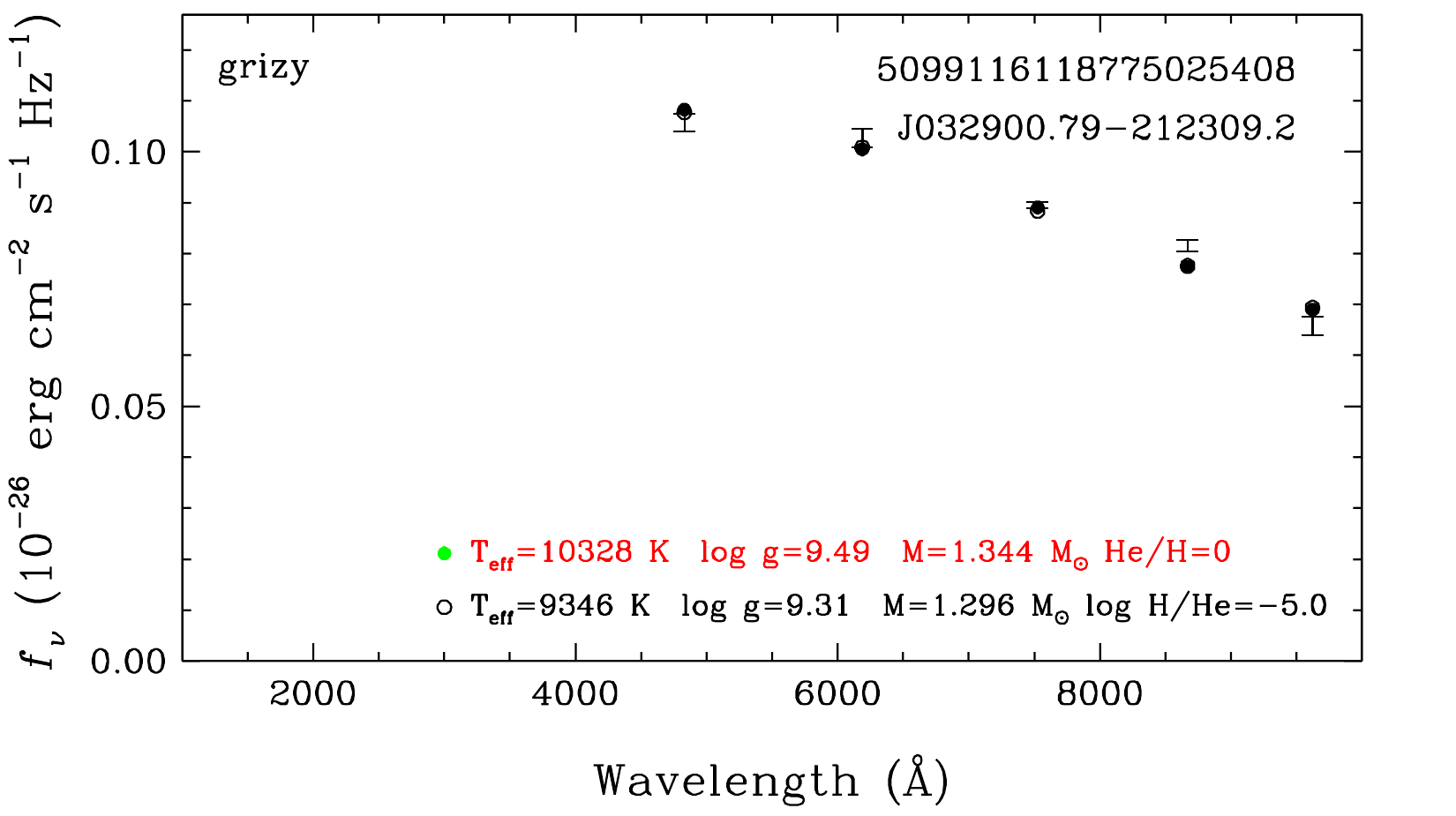}
\includegraphics[width=2.3in, clip=true, trim=0.9in 6.3in 1in 1.0in]{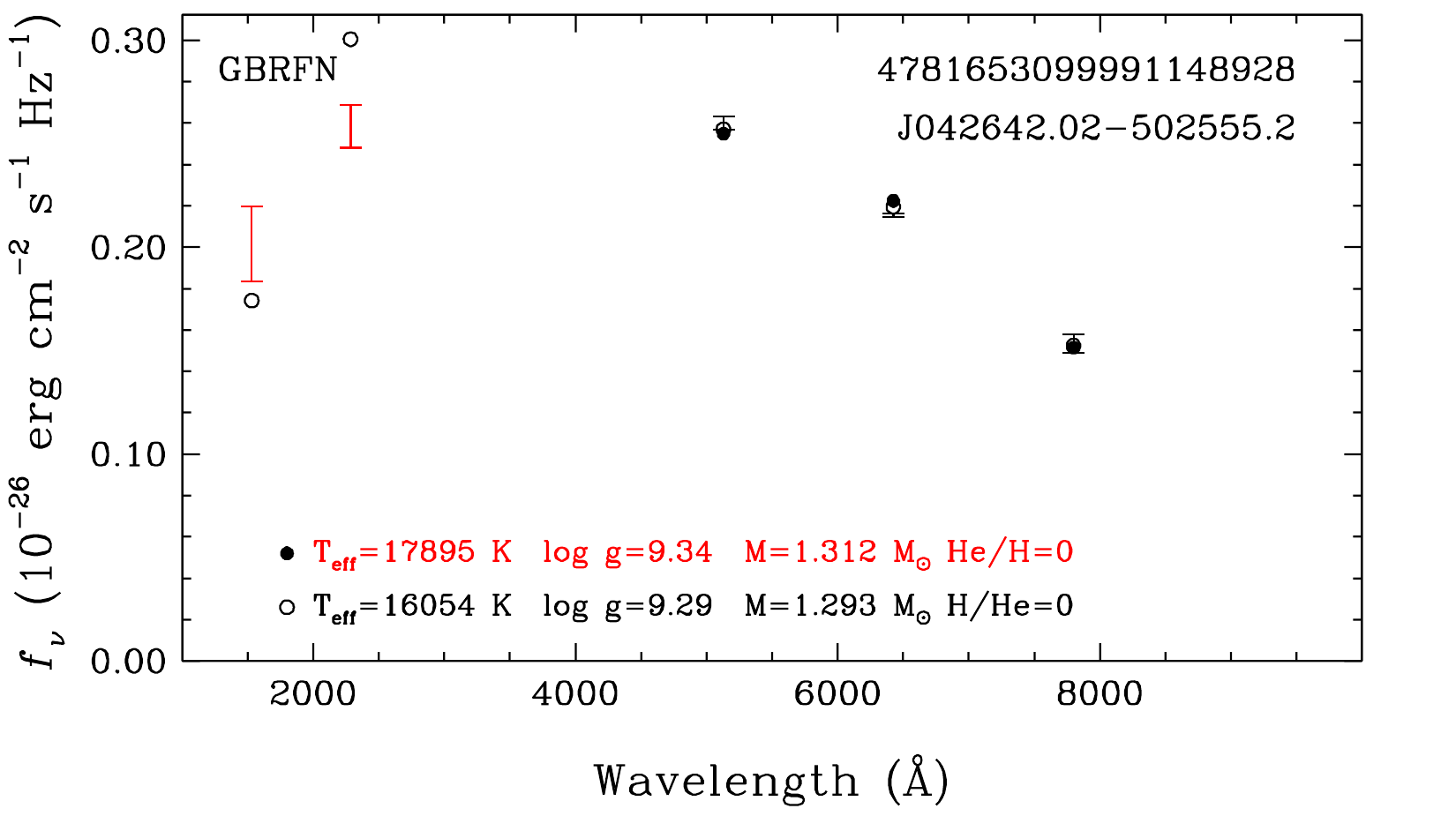}
\includegraphics[width=2.3in, clip=true, trim=0.9in 6.3in 1in 1.0in]{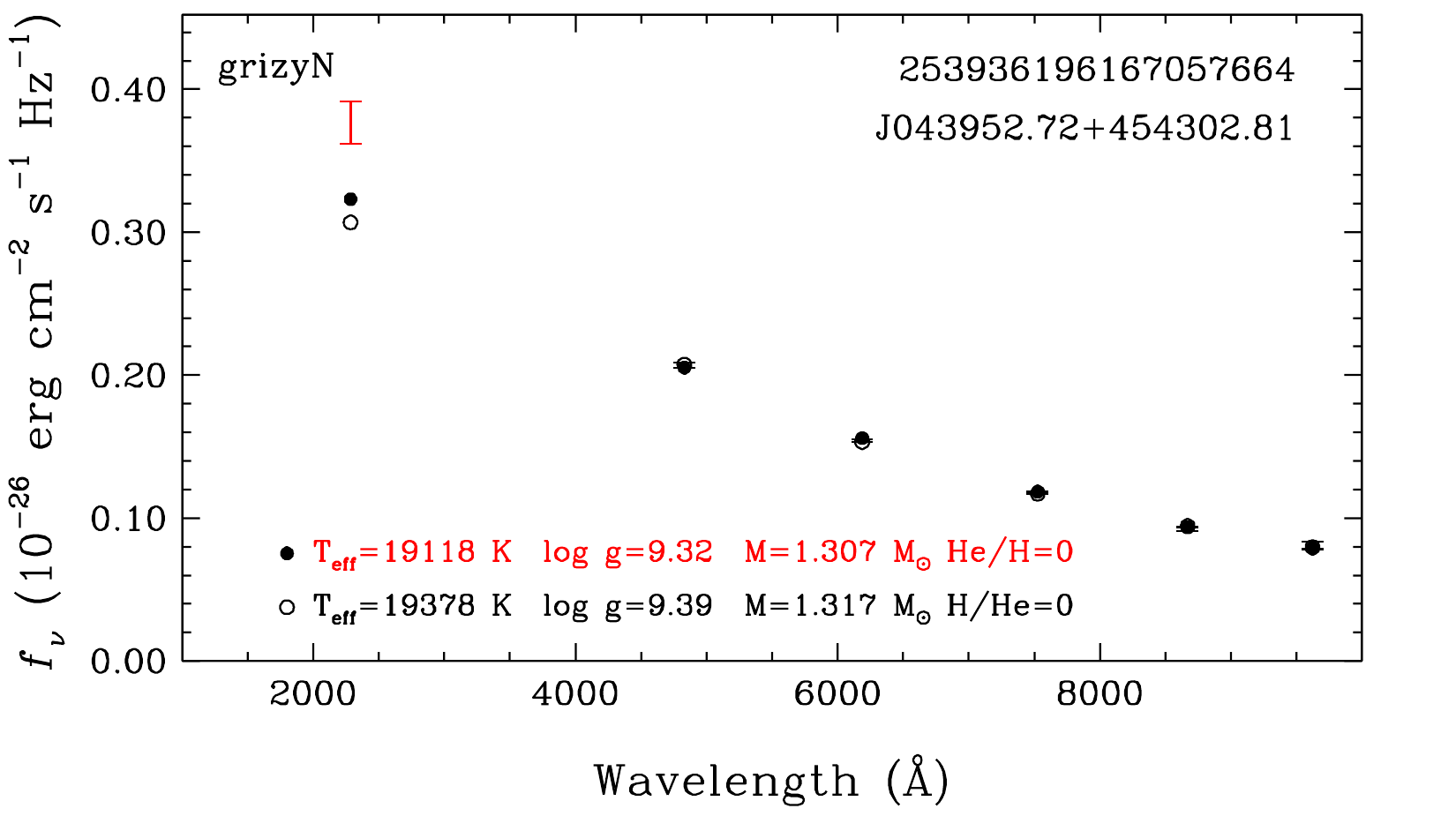}
\includegraphics[width=2.3in, clip=true, trim=0.9in 6.3in 1in 1.0in]{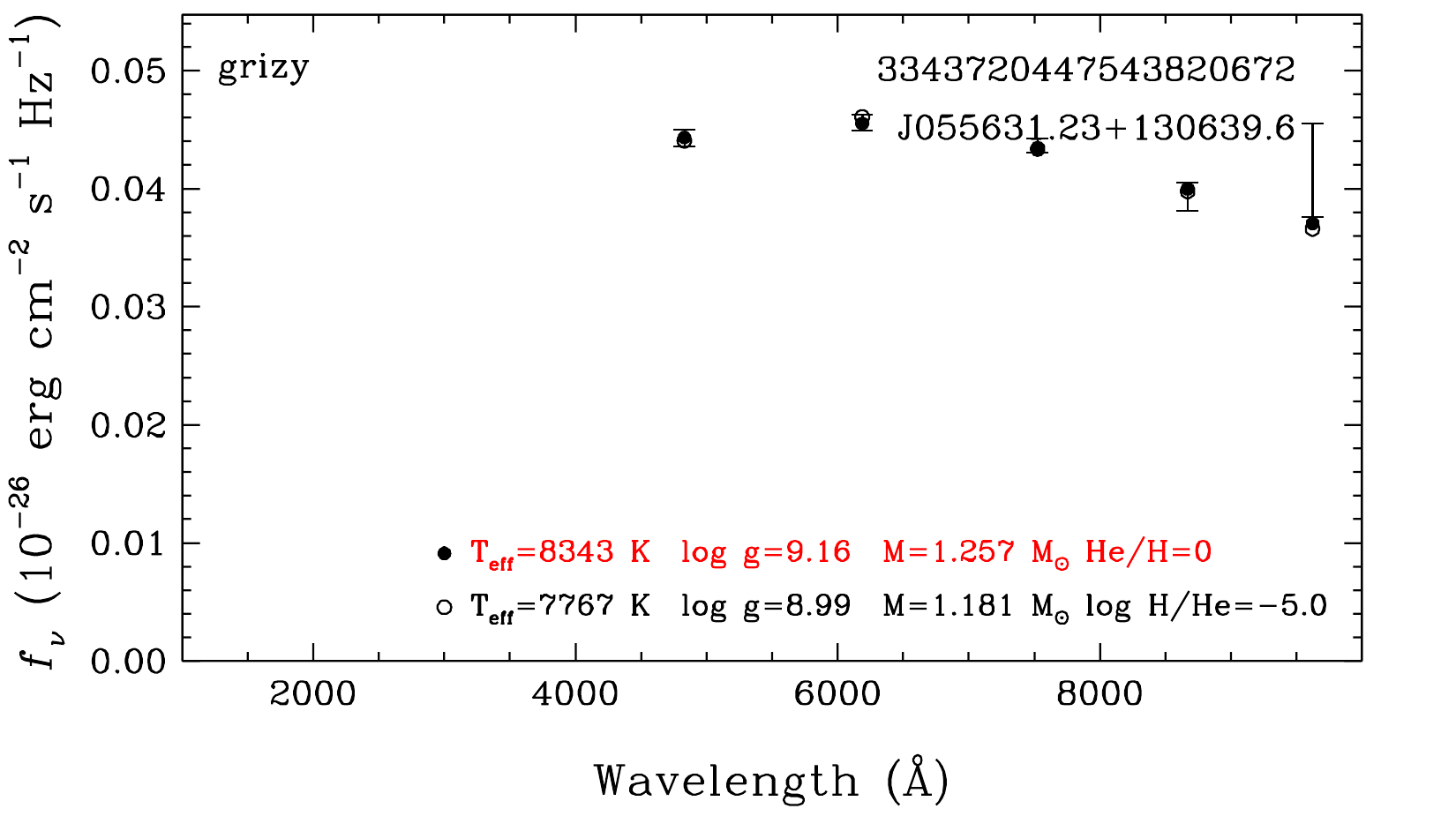}
\includegraphics[width=2.3in, clip=true, trim=0.9in 6.3in 1in 1.0in]{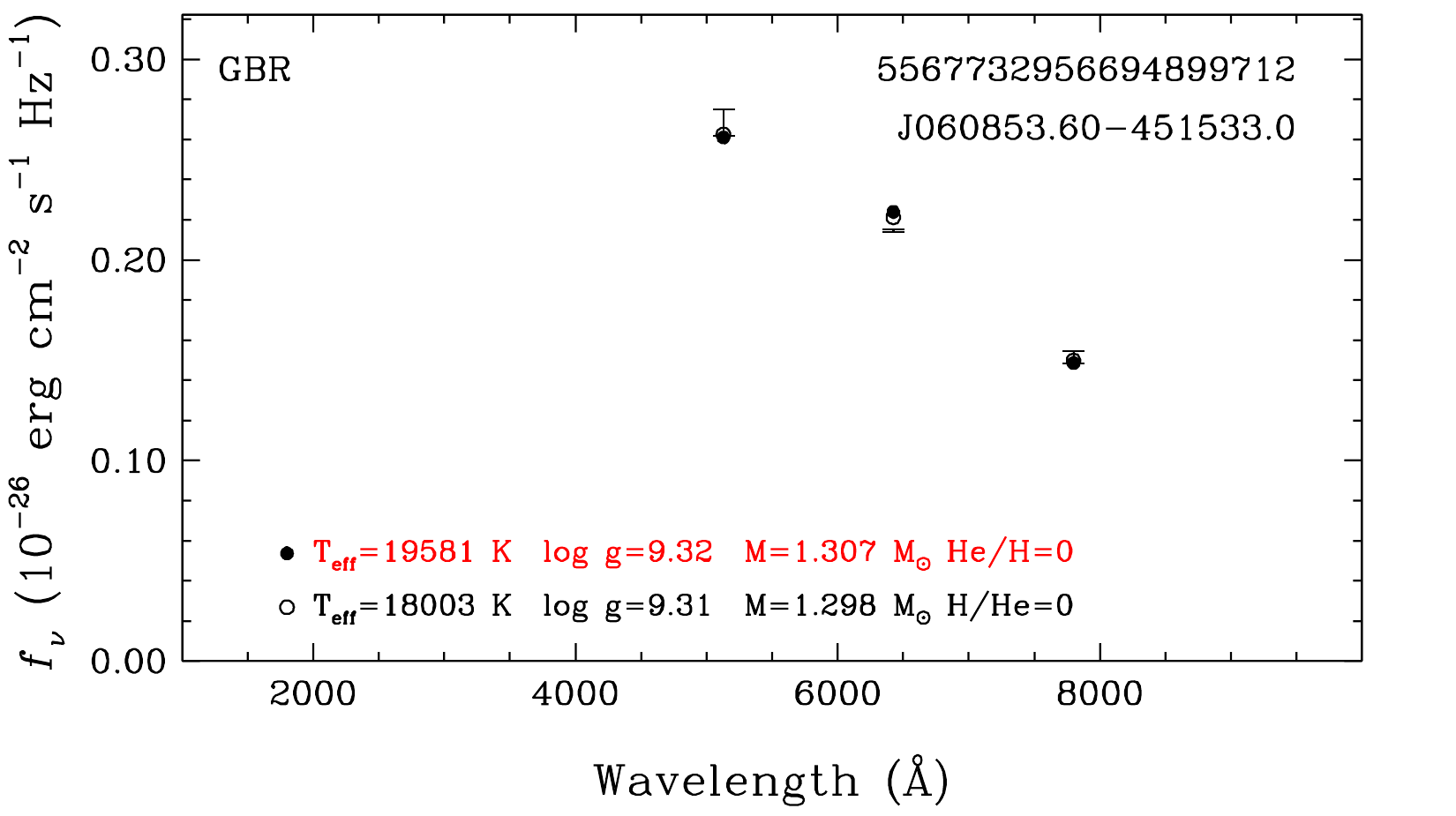}
\includegraphics[width=2.3in, clip=true, trim=0.9in 6.3in 1in 1.0in]{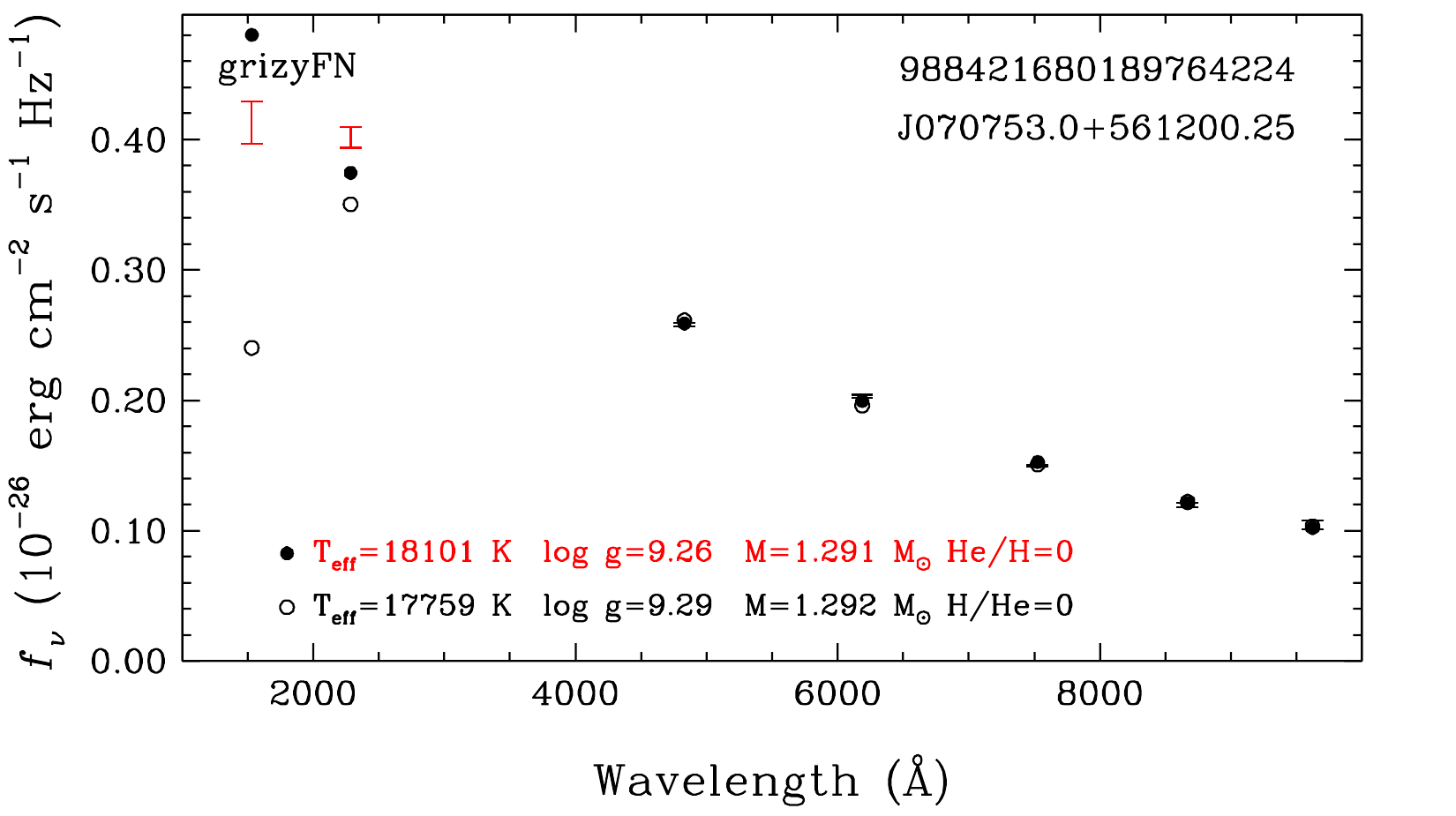}
\includegraphics[width=2.3in, clip=true, trim=0.9in 6.3in 1in 1.0in]{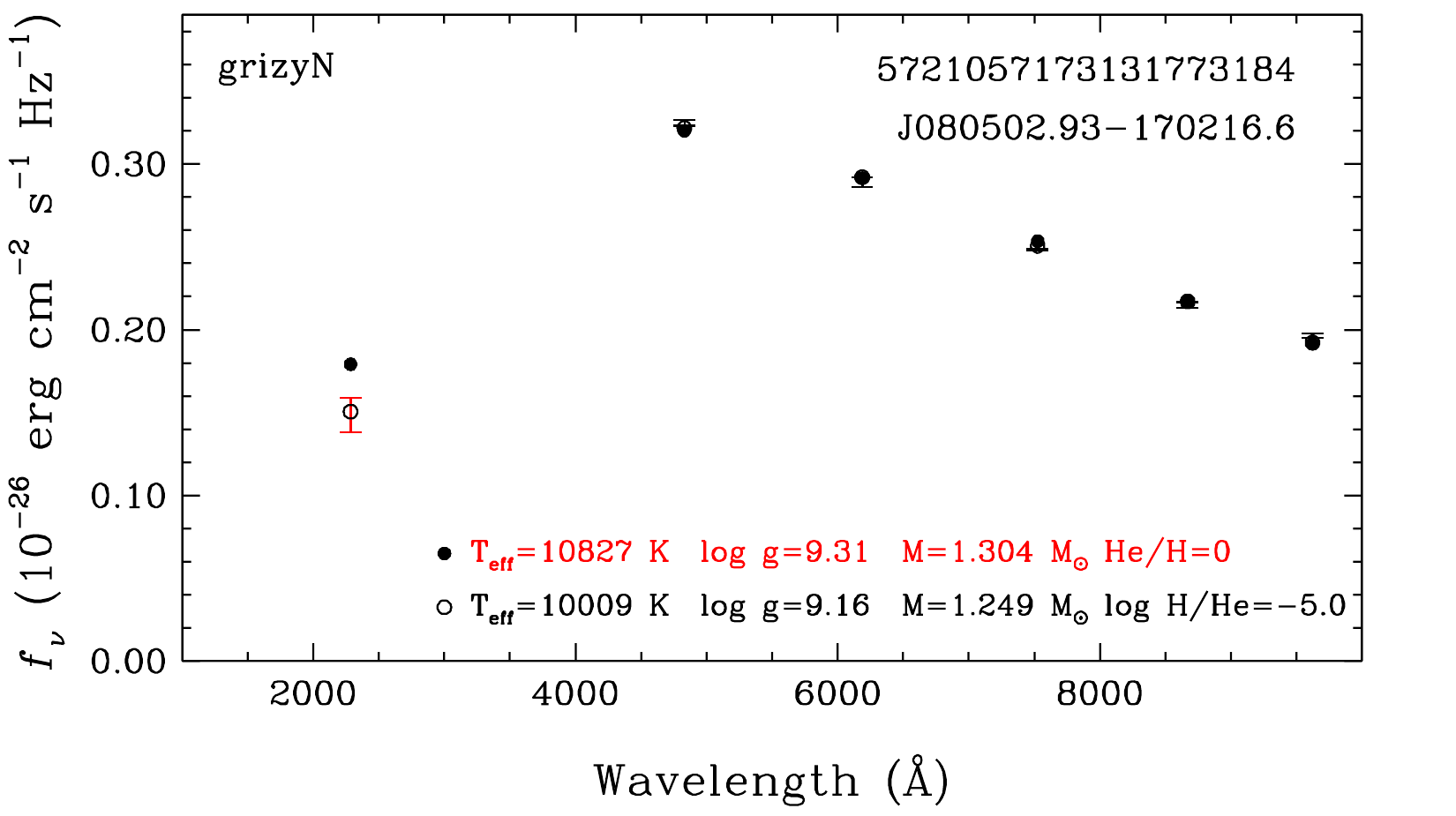}
\includegraphics[width=2.3in, clip=true, trim=0.9in 6.3in 1in 1.0in]{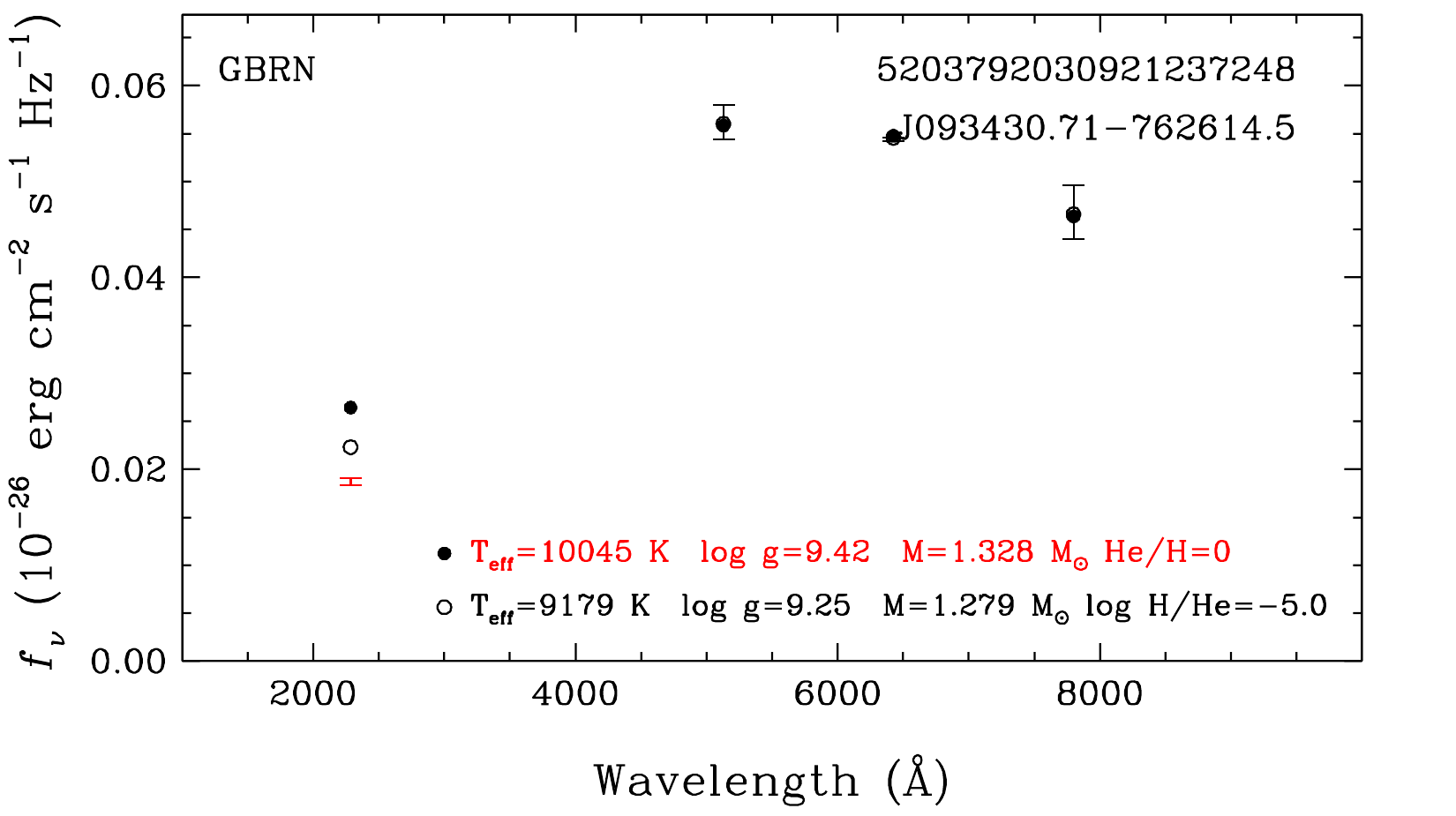}
\includegraphics[width=2.3in, clip=true, trim=0.9in 6.3in 1in 1.0in]{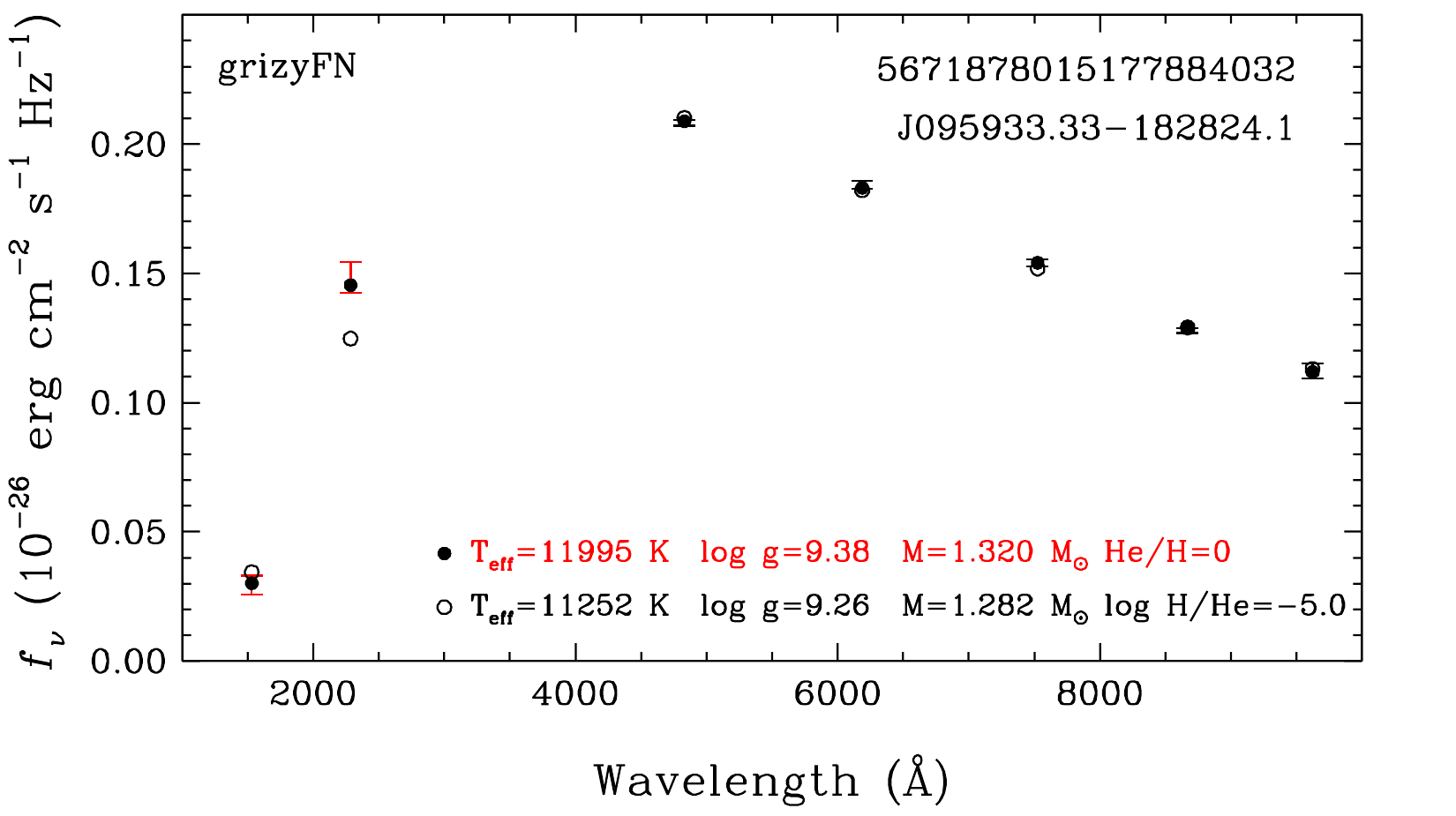}
\includegraphics[width=2.3in, clip=true, trim=0.9in 6.3in 1in 1.0in]{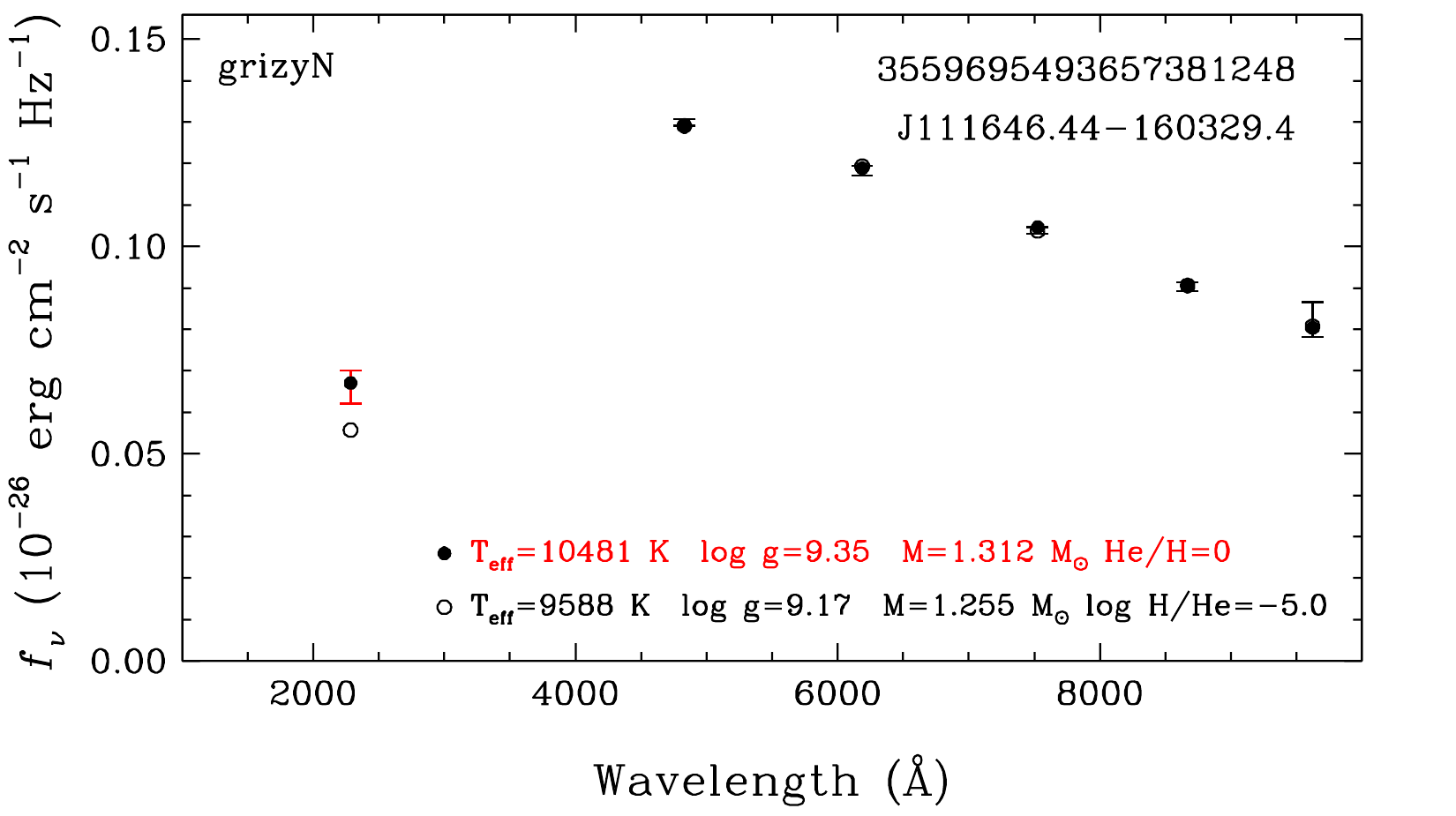}
\includegraphics[width=2.3in, clip=true, trim=0.9in 6.3in 1in 1.0in]{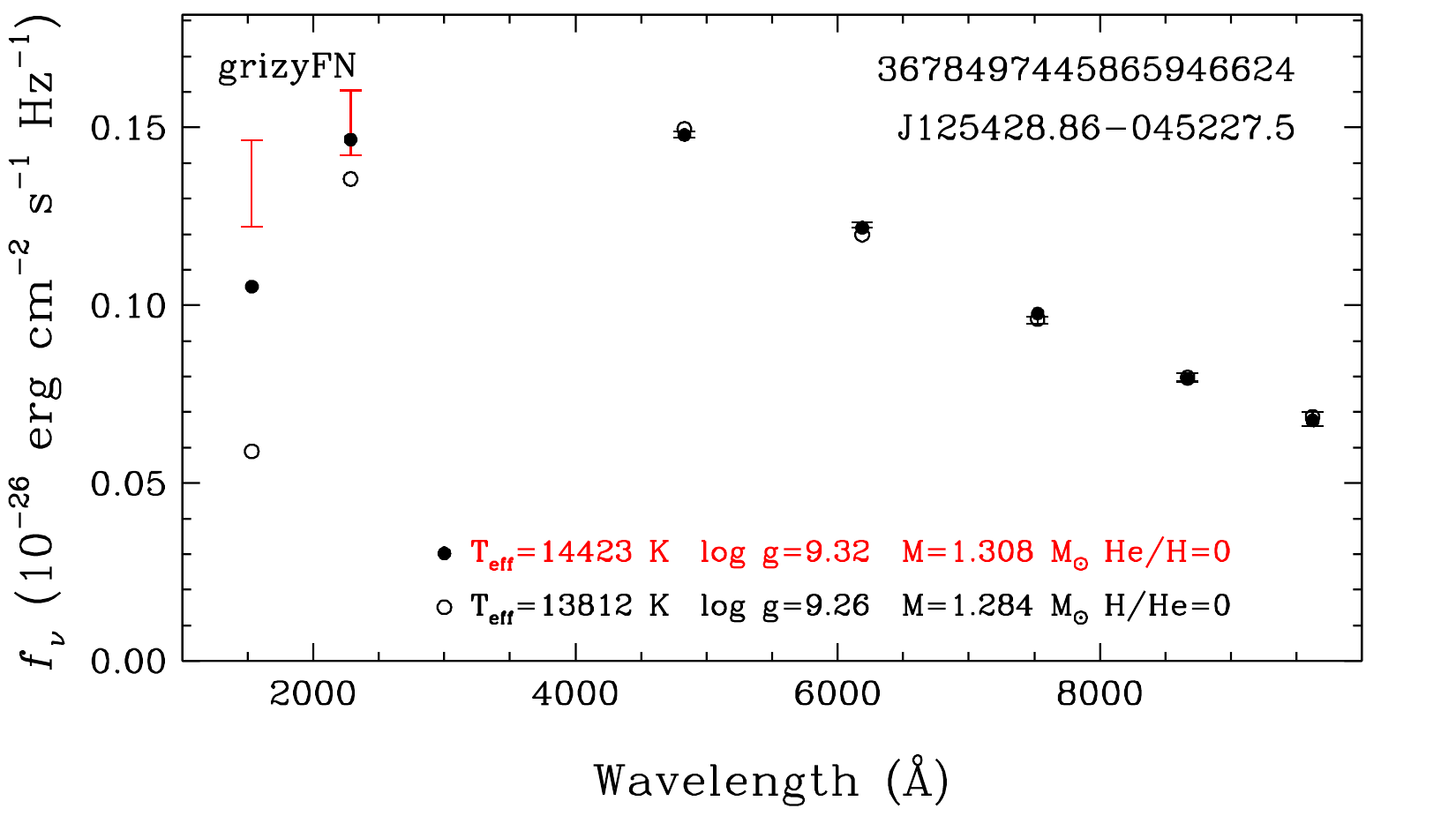}
\includegraphics[width=2.3in, clip=true, trim=0.9in 6.3in 1in 1.0in]{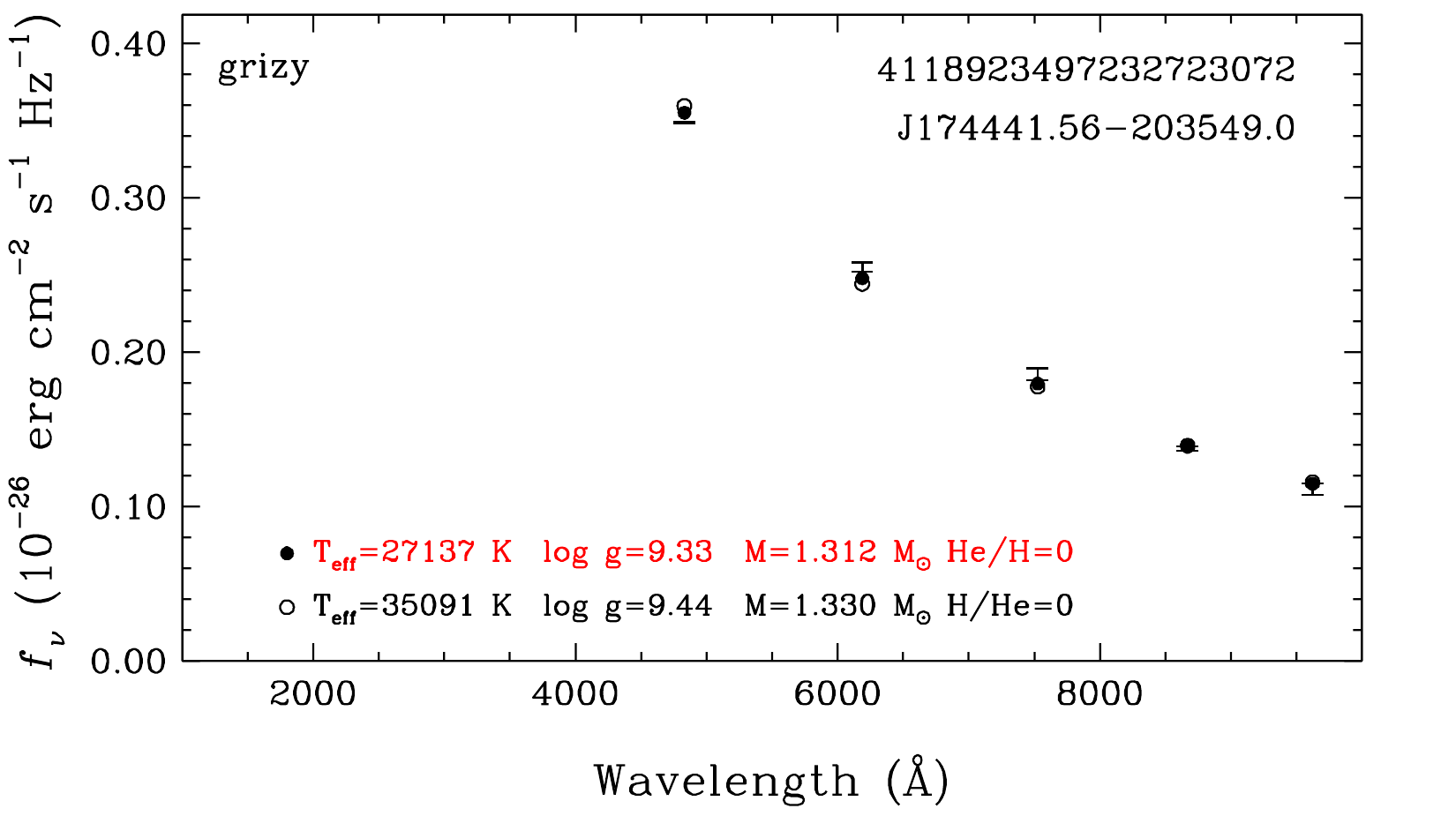}
\includegraphics[width=2.3in, clip=true, trim=0.9in 6.3in 1in 1.0in]{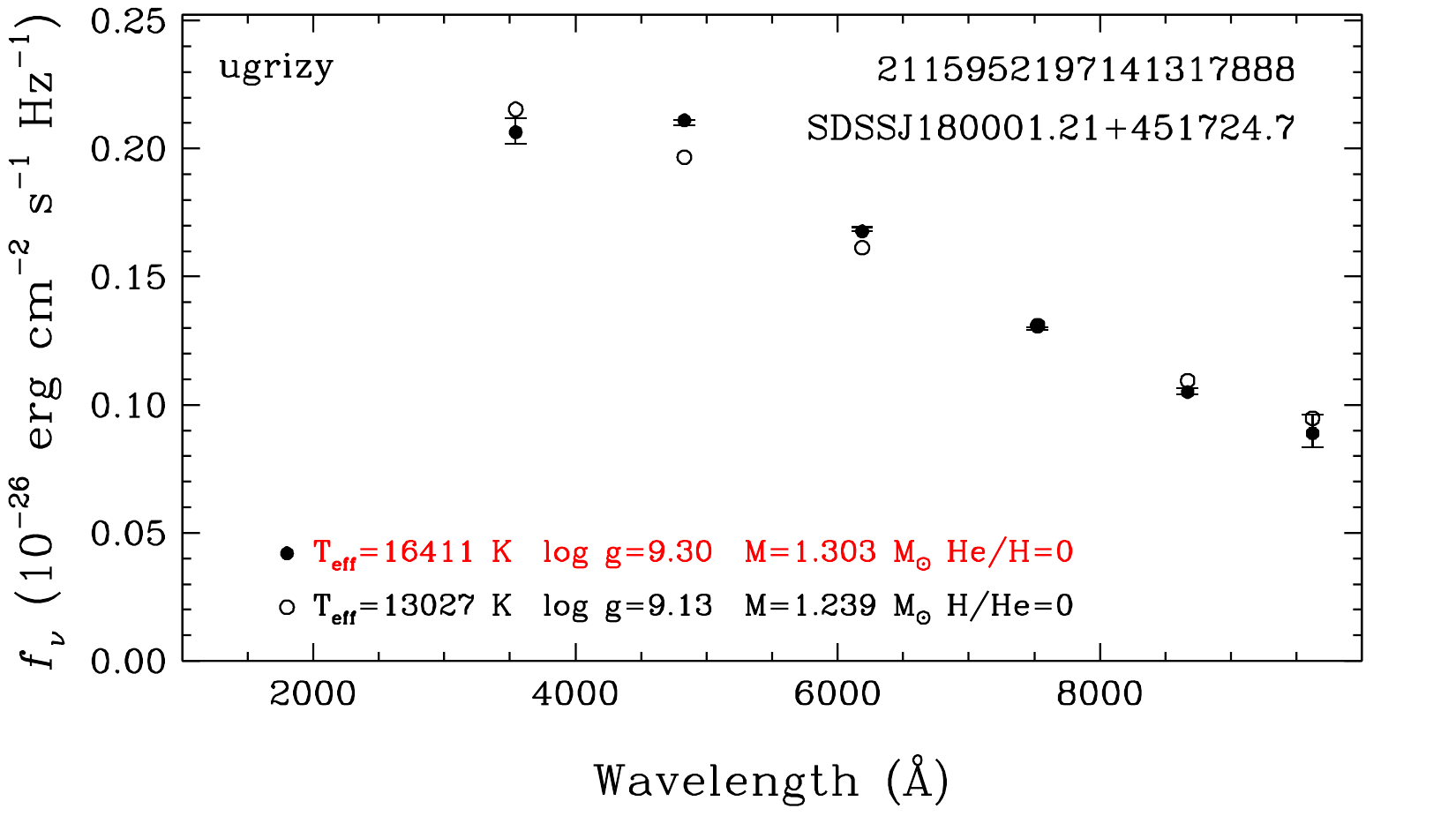}
\includegraphics[width=2.3in, clip=true, trim=0.9in 6.3in 1in 1.0in]{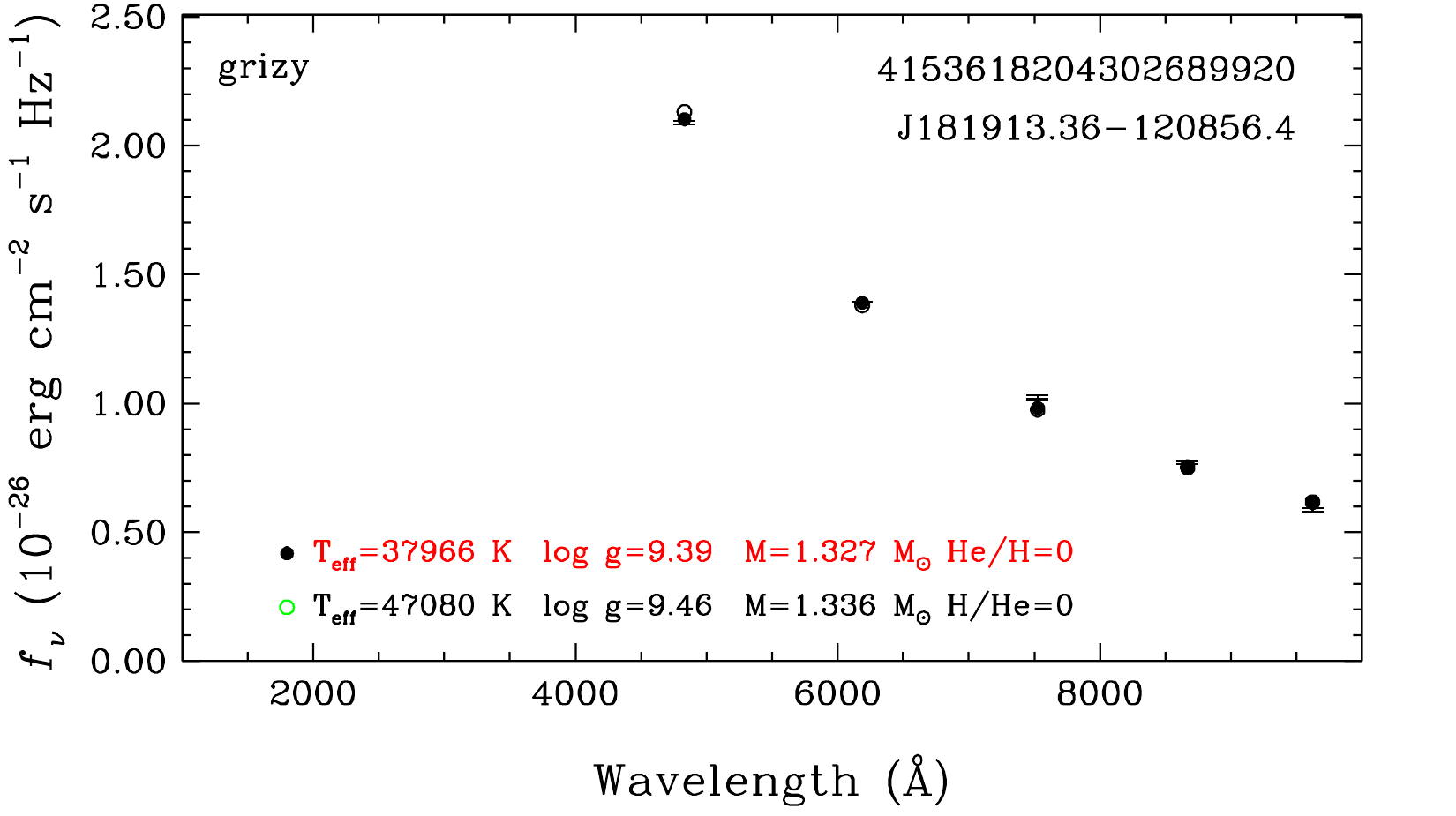}
\caption{Model atmosphere fits to 15 ultramassive white dwarf candidates without follow-up spectroscopy.
The symbols are the same as in Figure \ref{fig1727}. The best-fitting parameters are for CO cores.}
\label{figphot}
\end{figure*}

\subsubsection{Magnetic White Dwarfs and Others}
\label{mag}

The left panels in Figure \ref{figdc} show our model atmosphere analysis for the magnetic DA white dwarf J2211+1136, which
shows a weak H$\alpha$ line. Since we do not model the magnetic field structure, our best-fitting pure H solution does not match the H$\alpha$ line
profile. Regardless, this is clearly a DA white dwarf, and our analysis indicates a mass of $1.310 \pm 0.008~M_{\odot}$ for a CO core.

IPHAS J190132.77+145807.6 (middle panels in Figure \ref{figdc}) was classified as a DC white dwarf by \citet{deacon09} since it does not show
any significant features in its optical spectrum. However, it is too hot to be truly a DC white dwarf, as H or He lines should
be visible at a surface temperature near 30,000 K. Hence, it must be a magnetic white dwarf, either a DAH or a DBH.
Regardless of the atmospheric composition, J1901+1458 is clearly a very massive white dwarf with
$M=1.319 \pm 0.004~M_{\odot}$ (pure H solution) or $1.330 \pm 0.003~M_{\odot}$ (pure He solution) for a CO core.
 
Similarly, J2255+0710 (right panels) has a DC-like spectrum, but there are some weak features, including a broad
unidentified absorption feature near 4130 \AA. Hence, J2255+0710 is likely magnetic as well. Depending on the
atmospheric composition, its mass is in the range 1.22-1.30 $M_{\odot}$ for a CO core.

There are two additional targets with follow-up spectroscopy available in the literature. The first one,
J0103$-$0522, is DAH: white dwarf included in the 40 pc sample of
\citet{tremblay20}. The remaining target, J1832+0856, is a rapidly rotating DBA white dwarf \citep{pshirkov20}.
Our analysis using pure He atmosphere models for this object results in $T_{\rm eff} = 34200 \pm 1020$ K and
$M = 1.319 \pm 0.004~M_{\odot}$, assuming a CO core.

\begin{table*}
\centering
\caption{Physical Parameteres of the Photometry Only Sample assuming ONe or CO cores. All solutions
above 1.29 $M_{\odot}$ for ONe core models and above 1.334 $M_{\odot}$ for CO core models are extrapolated..}
\begin{tabular}{cccccccc}
\hline
              &                    &                       &       ONe core & ONe  core & CO core & CO core\\
Object  & Composition &  $T_{\rm eff}$ &  Mass &  Cooling Age & Mass & Cooling Age\\
        &             &    (K)         & ($M_{\odot}$) &   (Gyr)  & ($M_{\odot}$) & (Gyr) \\
\hline
J004917.14-252556.81  & H      &  13020 $\pm$ 460 & 1.263 $\pm$ 0.011 & 1.94 $\pm$ 0.08 & 1.312 $\pm$ 0.010 & 1.72 $\pm$ 0.09 \\
\dots                               & He     &  12260 $\pm$ 390 & 1.236 $\pm$ 0.011 & 2.06 $\pm$ 0.08 & 1.278 $\pm$ 0.012 & 1.68 $\pm$ 0.07\\
J032900.79-212309.24  & H      &  10330 $\pm$ 290 & 1.305 $\pm$ 0.010 & 2.32 $\pm$ 0.06 & 1.344 $\pm$ 0.008 & 1.87 $\pm$ 0.09\\
\dots  & $\log{\rm H/He}=-5$ &   9350 $\pm$ 250 & 1.256 $\pm$ 0.011 & 2.74 $\pm$ 0.07 & 1.296 $\pm$ 0.010 & 2.15 $\pm$ 0.10\\
J042642.02-502555.21  & H      &  17900 $\pm$ 1570 & 1.264 $\pm$ 0.019 & 1.30 $\pm$ 0.16 & 1.312 $\pm$ 0.016 & 1.08 $\pm$ 0.16\\
\dots  & He     &  16050 $\pm$ 1680 & 1.253 $\pm$ 0.031 & 1.51 $\pm$ 0.21 & 1.293 $\pm$ 0.028 & 1.15 $\pm$ 0.19\\
J043952.72+454302.81  & H      &  19120 $\pm$ 630 & 1.258 $\pm$ 0.008 & 1.18 $\pm$ 0.06 & 1.307 $\pm$ 0.007 & 0.96 $\pm$ 0.06\\
\dots  & He     &  19380 $\pm$ 1300 & 1.283 $\pm$ 0.019 & 1.24 $\pm$ 0.11 & 1.317 $\pm$ 0.014 & 0.81 $\pm$ 0.10\\
J055631.17+130639.78  & H      &   8340 $\pm$ 260 & 1.207 $\pm$ 0.021 & 3.33 $\pm$ 0.12 & 1.257 $\pm$ 0.023 & 3.34 $\pm$ 0.18\\
\dots  & $\log{\rm H/He}=-5$ &   7770 $\pm$ 170 & 1.157 $\pm$ 0.029 & 3.53 $\pm$ 0.07 & 1.181 $\pm$ 0.029 & 3.33 $\pm$ 0.13\\
J060853.60-451533.03  & H      &  19580 $\pm$ 1910 & 1.258 $\pm$ 0.021 & 1.13 $\pm$ 0.16 & 1.307 $\pm$ 0.019 & 0.92 $\pm$ 0.17\\
\dots  & He &  18000 $\pm$ 2800 & 1.259 $\pm$ 0.047 & 1.31 $\pm$ 0.29 & 1.298 $\pm$ 0.040 & 0.96 $\pm$ 0.26\\
J070753.00+561200.25  & H      &  18100 $\pm$ 350 & 1.240 $\pm$ 0.005 & 1.23 $\pm$ 0.04 & 1.291 $\pm$ 0.005 & 1.06 $\pm$ 0.04\\
\dots  & He     &  17760 $\pm$ 580 & 1.252 $\pm$ 0.009 & 1.31 $\pm$ 0.06 & 1.292 $\pm$ 0.009 & 0.98 $\pm$ 0.05\\
J080502.93-170216.57  & H      &  10830 $\pm$ 110 & 1.254 $\pm$ 0.004 & 2.40 $\pm$ 0.03 & 1.304 $\pm$ 0.003 & 2.20 $\pm$ 0.03\\
\dots  & $\log{\rm H/He}=-5$ &  10010 $\pm$ 120 & 1.213 $\pm$ 0.004 & 2.70 $\pm$ 0.04 & 1.249 $\pm$ 0.006 & 2.23 $\pm$ 0.04\\
J093430.71-762614.48  & H      &  10050 $\pm$ 1350 & 1.284 $\pm$ 0.055 & 2.47 $\pm$ 0.35 & 1.328 $\pm$ 0.047 & 2.11 $\pm$ 0.50\\
\dots  & $\log{\rm H/He}=-5$ &   9180 $\pm$ 1050 & 1.238 $\pm$ 0.052 & 2.86 $\pm$ 0.33 & 1.279 $\pm$ 0.051 & 2.32 $\pm$ 0.46\\
J095933.33-182824.16  & H      &  12000 $\pm$ 180 & 1.273 $\pm$ 0.005 & 2.12 $\pm$ 0.03 & 1.320 $\pm$ 0.004 & 1.83 $\pm$ 0.04\\
\dots  &    $\log{\rm H/He}=-5$ &  11250 $\pm$ 190 & 1.241 $\pm$ 0.007 & 2.31 $\pm$ 0.05 & 1.282 $\pm$ 0.007 & 1.85 $\pm$ 0.04\\
J111646.44-160329.42  & H      &  10480 $\pm$ 170 & 1.264 $\pm$ 0.007 & 2.45 $\pm$ 0.05 & 1.312 $\pm$ 0.006 & 2.21 $\pm$ 0.07\\
\dots  & $\log{\rm H/He}=-5$ &   9590 $\pm$ 160 & 1.218 $\pm$ 0.007 & 2.82 $\pm$ 0.05 & 1.255 $\pm$ 0.010 & 2.33 $\pm$ 0.05\\
J125428.86-045227.48  & H      &  14420 $\pm$ 390 & 1.258 $\pm$ 0.008 & 1.71 $\pm$ 0.06 & 1.308 $\pm$ 0.007 & 1.52 $\pm$ 0.06\\
\dots  & He     &  13810 $\pm$ 310 & 1.243 $\pm$ 0.009 & 1.79 $\pm$ 0.05 & 1.284 $\pm$ 0.009 & 1.43 $\pm$ 0.04\\
J174441.56-203549.05  & H      &  27140 $\pm$ 890 & 1.271 $\pm$ 0.008 & 0.65 $\pm$ 0.06 & 1.312 $\pm$ 0.008 & 0.43 $\pm$ 0.04\\
\dots  & He     &  35090 $\pm$ 1410 & 1.317 $\pm$ 0.009 & 0.45 $\pm$ 0.05 & 1.330 $\pm$ 0.007 & 0.18 $\pm$ 0.03\\
J180001.21+451724.7 & H      &  16410 $\pm$ 290 & 1.253 $\pm$ 0.003 & 1.44 $\pm$ 0.03 & 1.303 $\pm$ 0.004 & 1.26 $\pm$ 0.04\\
\dots  & He     &  13030 $\pm$ 180 & 1.206 $\pm$ 0.004 & 1.85 $\pm$ 0.04 & 1.239 $\pm$ 0.006 & 1.57 $\pm$ 0.03\\
J181913.36-120856.44  & H      &  37970 $\pm$ 1940 & 1.305 $\pm$ 0.009 & 0.37 $\pm$ 0.03 & 1.327 $\pm$ 0.006 & 0.12 $\pm$ 0.03\\
\dots  & He     &  47080 $\pm$ 1000 & 1.327 $\pm$ 0.006 & 0.29 $\pm$ 0.01 & 1.336 $\pm$ 0.004 & 0.02 $\pm$ 0.01\\
\hline
\end{tabular}
\end{table*}

\subsection{Objects with Unknown Spectral Types}
\label{utype}

Figure \ref{figphot} shows the spectral energy distributions and our model atmosphere fits to 15 ultramassive white
dwarfs with no follow-up spectroscopy available in the literature. Even though the presence or lack of a Balmer
jump between the UV and optical filters can be used to distinguish between the H- and He-atmosphere
solutions, we refrain from assigning composition based on photometry alone. However, we note that there are several
objects with spectral energy distributions that are best explained by H dominated atmospheres: J0049$-$2525, 
J0707+5612, J0959$-$1828, J1116$-$1603, J1254$-$0452, and J1800+4517 have SDSS $u$ or GALEX UV data
that clearly favor the pure H atmosphere solution. On the other hand, J0426$-$5025 and J0805$-$1702 have UV photometry
that favor He-dominated atmospheres. 

We provide both H- and He-dominated fits for each object in this figure, and list the best-fitting parameters for each
composition in Table 3. All of these targets have  $M\leq1.344~M_{\odot}$ assuming CO core composition.
Note that two objects, J0556+1306 and J0707+5612, have mass estimates below 1.3 $M_{\odot}$ for
the CO-core solutions. These two targets were selected as $M\geq 1.3~M_{\odot}$ white dwarfs based on Gaia photometry.
However, our improved fits using Gaia EDR3 parallax and Pan-STARRS photometry indicate masses slightly below
this limit.  

\section{Discussion}

\subsection{The Most Massive White Dwarf in the Solar Neighborhood}

By performing a detailed model atmosphere analysis of the MWDD 100 pc sample, we tried to identify the most massive
white dwarfs in the solar neighborhood. Among the 10 objects with follow-up optical spectroscopy available, 
there are two objects with masses above 1.334 $M_{\odot}$, the highest mass CO core model currently
available. Hence, their mass estimates are extrapolated, and therefore uncertain. With that caveat in mind,
the DA white dwarf J1329+2549 is currently the most massive white dwarf known in the solar neighborhood
with well constrained atmospheric parameters and a mass of $1.351 \pm 0.006 M_{\odot}$.

\citet{kilic20} estimated a mass of $1.358 \pm 0.022~M_{\odot}$ for the DA white dwarf J1140+2322, which is also included
in our sample. However, the previous analysis on this object was based on the Gaia DR2 parallax measurement and also
limited to CO core models up to 1.2 $M_{\odot}$.
Based on evolutionary sequences extended up to 1.334 $M_{\odot}$ and the Gaia EDR3 parallax measurement, which
implies a slightly larger distance, we derive a mass of $1.336 \pm 0.006~M_{\odot}$ for J1140+2322, making it the second
most massive white dwarf in our spectroscopy sample.

The MWDD sample selection is optimized for creating a clean white dwarf sample rather than completeness. 
Searching the Gaia DR2 white dwarf catalog of \citet{gentile19} for $M\geq1.3 M_{\odot}$ objects within 100 pc
reveals 40 high probability candidates with masses up to 1.37 $M_{\odot}$ under the assumption of a pure H atmosphere and a CO core.
These mass estimates are based on Gaia DR2 photometry and parallaxes. Three of these systems did not make it into the clean
candidate selection in the MWDD. Of the remaining 37 targets, 19 are included in our sample and in Table 1, but the remaining
18 are excluded because they appear less massive or because of our color and magnitude selection avoiding the IR-faint white dwarf sequence.

Among the 18 excluded objects, six have follow-up spectroscopy available in the literature. Three of these, Gaia DR2
2533306985471073920, 601566038739612160, 1358301480583401728, are in the SDSS footprint, and additional analysis by
\citet{kilic20} using Gaia parallax and SDSS + Pan-STARRS photometry showed that their masses are below 1.3 $M_{\odot}$. 
Another object, SDSS J071816.41+373139.1, could be included in our list of ultramassive white dwarfs if it has a He-dominated
atmosphere. The best-fitting He-dominated atmosphere solution has $T_{\rm eff} = 40636 \pm 1506$ K and
$M = 1.316 \pm 0.008~M_{\odot}$ \citep{kilic20}. However, we do not expect to see a DC white dwarf
at such a high temperature, unless it is a magnetic DBH or DAH; its atmospheric composition is uncertain. The pure H solution
implies a mass below $1.3~M_{\odot}$. Hence, it is not surprising that it is excluded from our list of 25 ultramassive white dwarfs
presented in Table 1. Two additional objects, J1001+3903 and J1337+0001 \citep{harris01,gates04}, are spectroscopically confirmed
IR-faint (ultracool) white dwarfs. 

In summary, we are confident that we are not missing a large population of $M\geq1.3~M_{\odot}$ ultramassive white dwarfs
within the Gaia 100 pc sample. However, Gaia itself is missing a number of cool ultramassive white dwarfs, since such white
dwarfs disappear quickly below Gaia's $G=20$ mag detection limit \citep{bergeron19,kilic20}.

\subsection{Single Stars versus Mergers}

A large fraction of the ultramassive white dwarfs in our sample have likely formed as a result of binary mergers.
Based on population synthesis calculations, \citet{temmink20} suggest that around 30 to 50\% of single white dwarfs with
$M>0.9~M_{\odot}$ form through a binary merger. 

Merger populations can reveal themselves through their kinematics. Since merger
systems would be found in older populations that are kinematically heated up, they should show a higher velocity dispersion
compared to a population of single stars \citep{wegg12}. For example, studying the transverse velocity distribution of a large sample of
$M=0.8-1.3~M_{\odot}$ white dwarfs, \citet{cheng20} show that 20\% of the massive white dwarfs in their sample must come
from mergers. Similarly, \citet{kilic20} found that 10 of their 44 hot DA white dwarfs with $M > 1~M_{\odot}$ have transverse
velocities in excess of 50 \kms, indicating a merger origin. 

\begin{figure}
\centering
\includegraphics[width=3.2in, bb=18 144 592 718]{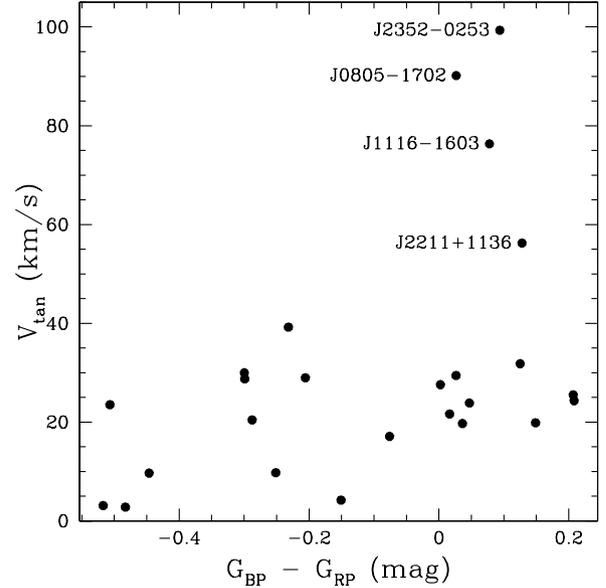}
\caption{Gaia colors (from DR2) and tangential velocities (from EDR3) of our sample of ultramassive white dwarf candidates presented in Table 1.The four velocity outliers are labelled.}
\label{figvtan}
\end{figure}

Figure \ref{figvtan} shows the Gaia colors and tangential velocities of our sample of 25 ultramassive
white dwarfs presented in Table 1. Excluding the $3\sigma$ outliers, the average tangential velocity of the sample
is $21 \pm 10$ \kms, which is consistent with a young disk population. There are four objects with transverse velocities (see Table 1)
that are significantly higher than expected, $>50$ \kms, for their cooling ages. These include the DA white dwarfs
J2211+1136 (which is also magnetic) and J2352$-$0253 (LHS 4033), and two additional objects without follow-up spectroscopy,
J0805$-$1702 and J1116$-$1603. Hence, these four ultramassive white dwarfs likely formed through mergers.

Merger populations can also reveal themselves through magnetism. A magnetic dynamo can be generated during a merger event
though differential rotation within a common-envelope or an accretion disk \citep{briggs15}. \citet{tout08} suggest that all highly
magnetic white dwarfs, either found as single stars or as components of magnetic cataclysmic variables, have a binary origin
\citep[but see][]{caiazzo20}. \citet{briggs15} further argue that binary mergers can explain both the observed incidence of
magnetism and the mass distribution of highly magnetic white dwarfs.

Interestingly, out of the 10 ultramassive white dwarfs with follow-up spectroscopy, 4 are likely magnetic (see section \ref{mag}).
As discussed above, one of these stars, J2211+1136, also displays a large transverse velocity, further providing evidence for
its merger origin. 

Merger populations can also reveal themselves through rapid rotation. Modeling the evolution of the remnants
of double white dwarf mergers, \citet{schwab21} estimate rotation periods of 10-20 min for most remnants on the white
dwarf cooling track. However, they find that the most massive white dwarfs with $M\geq1.2~M_{\odot}$ likely have shorter
rotation periods of 5-10 min. 

Only one of the ultramassive white dwarf candidates in our sample has a rotation measurement available.
\citet{pshirkov20} found a rotation period of 353 s for the DBA white dwarf J1832+0856, which is entirely consistent with
the simulations by \citet{schwab21}. Hence, J1832+0856 has likely formed through binary evolution as well.
Combining the results from these three different techniques shows that at least 8 of the 25 ultramassive white dwarfs
in our sample are likely to have a binary origin.

\subsection{Future Prospects}

There are several factors that contribute to the rarity of ultramassive white dwarfs in the solar neighborhood. The steep slope
of the initial-mass function \citep{salpeter55} means that their progenitor main-sequence stars are rare. In addition, given their
smaller radii, ultramassive white dwarfs are fainter than average, hence harder to find in magnitude limited surveys. The latent
heat from crystallization can keep these white dwarfs brighter for longer, but once they reach the Debye cooling regime,
they cool rapidly and disappear from view. Their numbers are significantly depleted for ages older than a few Gyr.

Gaia DR2 has provided us with a sample of 25 ultramassive white dwarf candidates with $M\sim1.3~M_{\odot}$ and within 100 pc.
With increased precision in parallax and photometry, future Gaia data releases may reveal additional ultramassive white dwarfs
in the solar neighborhood.

The Rubin Observatory's 10-year Legacy Survey of Space and Time (LSST) will find millions of white dwarfs, which will include
many ultramassive white dwarfs as well. An exciting prospect with these discoveries is that the time-series photometry from
the LSST can be used to measure the rotation periods of ultramassive white dwarfs, and to identify rapidly rotating systems
that likely formed through binary mergers. Such measurements can also constrain, as a function of mass, the fraction of single white dwarfs
that form through mergers. 

\section*{Acknowledgements}

We acknowledge enlightening discussions with our late colleague G.~Fontaine regarding the cooling sequences
discussed in this work. MK thanks E. Baron for useful discussions.

This work is supported in part by the NSF under grant AST-1906379, the NSERC Canada,
and by the Fund FRQ-NT (Qu\'ebec). S.B. acknowledges support from the Laboratory Directed
Research and Development program of Los Alamos National Laboratory under project number
20190624PRD2.

Based on observations obtained at the Gemini Observatory, which is operated by the Association of Universities for Research in Astronomy, Inc., under a cooperative agreement with the NSF on behalf of the Gemini partnership: the National Science Foundation (United States), National Research Council (Canada), CONICYT (Chile), Ministerio de Ciencia, Tecnolog\'{i}a e Innovaci\'{o}n Productiva (Argentina), Minist\'{e}rio da Ci\^{e}ncia, Tecnologia e Inova\c{c}\~{a}o (Brazil), and Korea Astronomy and Space Science Institute (Republic of Korea).

\section*{Data availability}

The data underlying this article are available in the MWDD at
http://www.montrealwhitedwarfdatabase.org and in the Gemini Observatory Archive at
https://archive.gemini.edu, and can be accessed with the program numbers
GN-2020A-DD-113 and GN-2020B-FT-107.

\input{ms.bbl}

\bsp
\label{lastpage}

\end{document}